\documentclass[twocolumn,eqsecnum,aps,prc,tightenlines,floats,floatfix]{revtex4}%%

\def\bea{\begin{eqnarray}}
\def\eea{\end{eqnarray}}

\def\sfrac#1#2{{\textstyle \frac{#1}{#2}}}

\newcommand{\braket}[2]{\langle #1|#2\rangle}

\def\be{\begin{equation}}
\def\ee{\end{equation}}
\def\ba{\begin{eqnarray}}
\def\ea{\end{eqnarray}}

\usepackage{graphics}
\usepackage{graphicx}
\usepackage{epsf} 
\usepackage{amsmath}
\usepackage{amssymb}
\begin{document}

% a4 form only
%\vspace{+2.2cm}

\phantom{0}
\vspace{-0.2in}
\hspace{5.5in}
\parbox{1.5in}{ \leftline{JLAB-THY-08-900}}

\vspace{-1in}%\parbox{1.5in}{ \vspace{-9.6in}}  % moves the preprint box down

\title
{\bf D-state effects in the electromagnetic $N \Delta$ transition}
\author{G. Ramalho$^{1,2}$, M.T. Pe\~na$^{2,3}$ and 
Franz Gross$^{1,4}$ 
\vspace{-0.1in}  }

\affiliation{
$^1$Thomas Jefferson National Accelerator Facility, Newport News, 
VA 23606 \vspace{-0.15in}}
\affiliation{
$^2$Centro de F{\'\i}sica Te\'orica de Part{\'\i}culas, 
Av.\ Rovisco Pais, 1049-001 Lisboa, Portugal \vspace{-0.15in}}
\affiliation{
$^3$Department of Physics, Instituto Superior T\'ecnico, 
Av.\ Rovisco Pais, 1049-001 Lisboa, Portugal \vspace{-0.15in}}
\affiliation{
$^4$College of William and Mary, Williamsburg, VA 23185}

\begin{abstract} 

We consider here a manifestly covariant quark model 
of the nucleon and the $\Delta$, where one quark is off-shell and the other 
two quarks form an on-shell  diquark pair.
Using this model, we have shown previously that 
the nucleon form factors and the dominant form factor
for the $\gamma N \to \Delta$ transition 
(the magnetic dipole (M1) form factor) 
can be well described by nucleon and 
$\Delta$ wave functions with S-state components only.
In this paper we show that non-vanishing results for the small electric (E2)
and Coulomb (C2) quadrupole form factors can be obtained if   
D-state components are added to the $\Delta$ valence quark wave function. 
We present a covariant definition of these components and
compute their contributions to the form factors.  
We find that these components
  cannot, by themselves,
   describe the data. 
    Explicit pion cloud contributions must also be 
added and these contributions
    dominate both the E2 and the C2 form factors.
By  parametrizing the pion cloud contribution for 
the transition electric and Coulomb form factors 
in terms of the neutron electric form factor, 
we estimate that the contributions of the $\Delta$ D-state
coupled to quark core spin of 3/2  is of the order of 1\%,
and the contributions of the $\Delta$ D-state coupled to 
quark core spin 1/2 is of the order of 4\%.
%\maketitle
\end{abstract}

\date{\today}

\phantom{0}
%\vspace{7.0in}
%\vspace{-6in}
\vspace*{0.9in}  % sets how far the title is below the preprint box
\maketitle

\section{Introduction}

Understanding the internal structure of the baryons
is both an experimental and a theoretical
challenge.
Experimentally, the main source of information has been
the electro- and photo- excitation of the nucleon,
which allows to parametrize the baryons internal structure in terms
of their electromagnetic form factors.
Very accurate  Jlab data \cite{Jones99,Gayou01,Punjabi05} 
exist nowadays for the nucleon elastic
form factors. Also, theoretical  models 
for the nucleon form factors are able to 
describe this data well
\cite{Nucleon,Gross08b,Gross04,Hyde-Wright04,Arrington06,Perdrisat07}.
The next step is the description of the 
nucleon excitations, starting with 
the $\Delta$ resonance. 
In the recent years new precise data have been collected 
from MAMI \cite{MAMI,Stave06a},
LEGS \cite{LEGS}, MIT-Bates \cite{Bates}
and Jlab \cite{CLAS02,CLAS06}
in the region $Q^2 \le 6$ GeV$^2$
($q^2=-Q^2$ is the squared  momentum transfer). 
The  $N \Delta$ electromagnetic transition 
($\gamma N \to \Delta$) has a simple 
interpretation in terms of the valence quark structure:
the $\Delta$ results from  a spin flip of a single quark
in the nucleon.
It is then understandable that the
magnetic dipole multipole M1 dominates 
the transition for low $Q^2$, 
and that the electric E2 and the Coulomb C2 quadrupoles 
give only small contributions, of the order of few percent.
For large $Q^2$ however, according to perturbative 
QCD (pQCD) \cite{Carlson,Sterman97},  
equally important contributions from M1 and E2 are to be expected,
but the scale for the outset of that regime is not yet known exactly.  

Several theoretical descriptions 
have been proposed for low, intermediate, 
as well as for the large transfer momentum $Q^2$ regions.
These descriptions involve two ingredients:
the valence quark and the non-valence degrees of freedom.
The non-valence degrees of freedom 
are essentially the sea quark contributions  
which represent quark-antiquark states, and are usually called
meson cloud effects.
Due to its pseudoscalar character 
and its low mass, chiral symmetry assigns a special role to the pion
\cite{Bernstein03,Bernstein07}, and
pion cloud effects   
are therefore expected to contribute significantly 
to the baryon excitations.
At low momentum transfer
Effective Field Theory models based on chiral symmetry
and perturbation theory ($\chi$PT)
\cite{Pascalutsa06a,Gail06,Ramirez06,Arndt04},
with nucleon, $\Delta$ and  pion 
degrees of freedom, and 
no internal structure considered, work well.
Effective Field Theories (EFT) 
describe the pion cloud 
effects at low momenta,  but have 
a limited range of application, $Q^2 < 0.25$ GeV$^2$.
At low $Q^2$, the large $N_c$ limit 
\cite{Pascalutsa07a,Dominguez07} 
can be used to establish the main $Q^2$ dependence 
of the form factors and derive relations 
between the nucleon and the $N\Delta$ form factors 
\cite{Pascalutsa07a}.
At large $Q^2$ models within
pQCD \cite{Carlson,Idilbi04} with quarks and gluons as 
degrees of the freedom, can be applied.
As for the intermediate momentum region, it may be appropriately featured by 
constituent quarks models
\cite{Giannini07,Bernstein07,Isgur82,Capstick95,Capstick90a,Capstick92,
Riska,DeSanctis04,Donoghue75,Warns90,Bijker94,Dong06,Yu08,
Keiner96,Gorchtein04}, 
and models based on hadronic 
degrees of freedom, as the so-called dynamical models (DM) 
\cite{SatoLee,SatoLeeII,Diaz06a,Kamalov99,Kamalov01,Pascalutsa04}.
Quark models with mixed coupling with 
pion fields have also been proposed 
\cite{Giannini07,Bernstein07,Kaelbermann83,Fiolhais96,Amoreira00,
Bermuth88,Lu97,Faessler06,BuchmannEtAl,LiRiska}.
In the intermediate regime
results from Vector Meson Dominance models
\cite{Vereshkov07},
QCD sum rules  \cite{Braun06,Rohrwild07} and 
Global Parton Distributions (GDPs) 
\cite{Stoler,Pascalutsa06c,Carlson08}, 
have been presented as well.
Finally, precise calculations are recently emerging 
from Lattice QCD calculations
\cite{Leinweber93,Alexandrou05,Alexandrou04,Alexandrou07}.
For a %comprehensive 
review of the
state-of-the-art in experiments and theory see 
Refs.~\cite{Bernstein07,Burkert04,Pascalutsa06b,Stave08}.

There is at present a strong motivation to pursue an
interplay between dynamical models and constituent quark models 
\cite{Giannini07,Bernstein07,Diaz06a,Burkert04}.
On one side, constituent quark models underestimate the result of 
the transition form factors, 
when not combined with explicit pion degrees of freedom 
\cite{Giannini07,Diaz06a,Pascalutsa06b,NDelta,Burkert04}. 
On the other hand, dynamical models are
based on sets of equations 
coupling electromagnetic excitations 
to meson (de)excitations of the baryons, and
include pion cloud effects
naturally and non-perturbatively. Examples are the 
Sato and Lee (SL) \cite{SatoLee,SatoLeeII,Diaz06a}, the
Dubna-Mainz-Taipei
(DMT) \cite{Kamalov99,Kamalov01} and the (Dynamical) 
Utrecht-Ohio \cite{Pascalutsa04} models. 
Although very successful in the description of the 
the $\gamma N \to \Delta$ form factors, they need 
to assume an initial parametrization for the 
baryon transition vertex,  interpreted as 
the bare vertex, where no pion loop is taken.
In a less ad-hoc fashion, the  bare vertex should therefore be derived 
from a quark model \cite{Burkert04}. 
At the same time we can use the available models 
to extract the bare vertex.
This is the goal of the EBAC program 
\cite{EBAC,Diaz06a,Burkert04}.
Alternative descriptions of 
the pion cloud and its relation 
to the deformations of the baryons 
were also proposed by 
Buchmann {\it et.~al.} 
\cite{BuchmannEtAl,Buchmann01,Buchmann04,Buchmann07a}. 
For an updated review of the dynamical models see 
Refs.~\cite{Diaz06a,Drechsel06,Tiator06}.

From the literature it is not clear 
which effects are due to the valence quarks, 
and which are related with the pion cloud, 
in particular for the E2 and C2 multipoles. 
There is a disagreement about those effects,
between models based on different formalisms, 
as the DM and EFT models,  and even between models based
on the same framework, as Effective Fields Theories 
\cite{Tiator06}.
Also in the experimental sector there are some ambiguities.
The form factors are extracted using 
multipole analysis based on Unitary 
Isobar Models (UIM) like MAID \cite{Drechsel07,MAID,MAID2}, 
SAID  \cite{Arndt95,Arndt96,Arndt03a,SAID2}
or Jlab/Yereven \cite{Aznauryan03,Aznauryan03b}, 
each leading to different results 
due to the differences in parametrizations 
of the background and resonance structures, 
even considering dynamical models 
\cite{SatoLee,SatoLeeII,Diaz06a,Kamalov99,Kamalov01,Pascalutsa04}
instead of UIM
\cite{Bernstein03,Stave08,Kamalov01,Diaz06a}.
The ambiguities involved in the interpretations of the data
are well illustrated by the 
differences between the CLAS results
\cite{CLAS06} and the MAID analysis of the same data
\cite{Drechsel07}, and also the recent preliminary
CLAS data analysis \cite{Stave08,Diaz06a,Smith07} based on
different models.
%A good example of the uncertainty involved in 
%the multipole extraction is the preliminary
%CLAS data \cite{Stave08,Diaz06a,Smith07}.
The results that we obtained here for $G_C^\ast$ in particular,
illustrate well the need to clarify these issues,
as we will discuss later in this paper.

In a previous work \cite{Nucleon,Gross08b,NDelta}
the spectator formalism \cite{Gross69,Gross92,GrossSurya} 
was applied to the nucleon and to the $\Delta$ baryons, 
considering only S-state wave functions. As shown in that work, 
with S-waves alone in the baryon
wave functions, only the dominant
of the three form factors for the  $\gamma N  \to \Delta$ transition
does not vanish.  Therefore
here we explore for the first time the effects 
of the D-states in the $\Delta$ 
wave function within that formalism, and show here that those 
components in the $\Delta$ wave function 
lead to non-zero contributions 
for the subleading form factors E2 and C2.
The origin of the D-wave states is well-known: 
in the pioneering work  of  Isgur-Karl \cite{Pascalutsa06b,Isgur82}
the baryons are described as a system 
of confined quarks, where a tensor color hyperfine interaction
is generated by one-gluon-exchange processes.
This tensor interaction leads 
to SU(6) symmetry breaking, and allows the transition 
from the ground S-state 
to an excited D-state.

For each of the three $N \Delta$ electromagnetic 
transition form factors, we  identified 
and separated the roles from the
different partial wave components.
While the magnetic dipole form factor 
$G_M^\ast$, the dominant contribution, 
is mainly due to the transition between 
the nucleon and  the S-state of the $\Delta$,
the electric quadrupole form factor $G_E^\ast$
proceeds through  the transition to a  D-state of the $\Delta$ 
corresponding to a three-quark core spin of 3/2.
Finally, the Coulomb quadrupole form factor $G_C^\ast$
becomes non-zero, only when the 
transition to a D-state of the $\Delta$  
corresponding to a three quark core spin 1/2 is switched on.
Nevertheless, and in agreement with other
quark models, we conclude that the valence 
quark effects are not sufficient to 
describe the E2 and C2 data \cite{Pascalutsa06a,Giannini07,Bernstein07}.

Additional mechanisms involving 
the sea quark states, mainly 
the pion cloud effects, 
are needed to fill the gap between the theory 
and the experimental data.
The systematic and consistent treatment 
of the pion cloud mechanisms is out 
of the scope of this work, which is focused on
the D-state effects,
but is planned for a future work.
In order to estimate the magnitude 
of the D-states we considered the
simple parametrization of 
the pion cloud in terms of the nucleon (neutron) 
electric form factor, with no additional parameters.
This parametrization 
was derived from the basic 
properties of the quark models 
(large $N_c$ limit and also SU(6) symmetry breaking)
and is limited in its 
range of application to low $Q^2$. Nevertheless, 
% ($Q^2 << 1$ GeV$^2$),
we need to include a pion cloud parametrization 
for a realistic estimate of  the weight of the D-wave components 
in the $\Delta$ wave function.

This paper is organized as follows: 
the formalism for the D-wave components of the 
$\Delta$ wave function is explained in Sec.~\ref{secWave Function},
the definitions of the form factors and other 
general results are introduced in Sec.~\ref{secFFgen},
the issue of gauge invariance and how it couples 
to the orthogonality of the initial and final 
state is dealt in Sec.~\ref{secEMcur},
the  formulas for the form factors within the valence quark model used here
are given in Sec.~\ref{secFFval}. 
In Sec.~\ref{secPionCloud} we discuss the 
the contributions of the sea quarks (pion cloud), 
and in Sec.~\ref{secResults} we present
the numerical results for representative models based 
on valence and sea quarks.
A discussion follows in Sec.~\ref{secResB}, and
conclusions are presented in Sec.~\ref{secConclusions}.

\section{Nucleon and $\Delta$ wave functions}
\label{secWave Function}

\begin{figure}[t]
\vspace{0.0cm}
\centerline{
\mbox{
\includegraphics[width=2in]{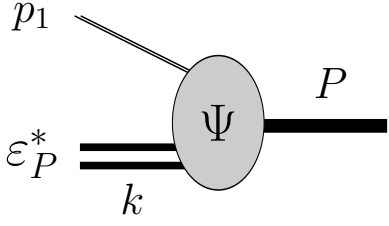}}}
\vspace{0.9cm}
\caption{\footnotesize{
Baryon quark-diquark wave function amplitude.}}
\label{figBaryonV}
\end{figure}

In the framework of the spectator theory 
\cite{Gross69,Gross92}
a baryon with four-momentum $P$ is taken to be a bound state of a
quark-diquark system, with relative four-momentum $k$,  and is described by a covariant amplitude $\Psi(P,k)$.
The diquark is taken to be on-mass-shell with an average mass $m_s$.
The 3-quark wave function amplitude depicted in the diagram of 
Fig.~\ref{figBaryonV}
%is then written as
%\be
%\Psi(P,k)=(m_q-\not\! p_1)^{-1} 
%\bra{k} \Gamma  \ket{P},
%\label{eqNvertex}
%\ee
has  S-wave components which, for the nucleon and the $\Delta$,  were already 
presented in Refs.\  \cite{Nucleon,NDelta}. 
Therefore they will be only very briefly reviewed here, where the main focus is
on the construction of D-wave components within the same underlying formalism.
In the following we will use $H$ to denote either the nucleon ($N$), with mass $m_H=m_N=m$, or the $\Delta$ with mass $m_H=m_\Delta=M$.

The antisymmetry for the color part of the baryon wave function  implies that,  for S and D-waves, the spin-isospin part of the wave function is 
symmetric.  This in turn implies that the diquark has positive parity.

\subsection{S-wave components of the Nucleon and $\Delta$ wave functions}

The S-wave part of the nucleon wave function has two components corresponding, respectively, to a diquark of spin 0-isospin 0 and a diquark of spin 1-isospin 1.  Labeling the polarization of the spin-1  diquark  by $\lambda$, these two terms for the nucleon amplitude shown in Fig.~\ref{figBaryonV} can be written
\bea
\Psi_{N\,\lambda_n}^S(P,k)&=& \frac{1}{\sqrt{2}} 
\left[\phi_I^0 u_{_N}(P,\lambda_n) - 
\phi_I^1 \varepsilon_{\lambda P}^{\alpha \ast}U_\alpha(P,\lambda_n) \right] \nonumber\\
&&\qquad \times\;\psi_N^S(P,k).
\label{eqPsiN}
\eea
The isospin states $\phi_I^{I_d}$ (with $I_d=0,1$ the diquark isospin) are 
respectively  $\phi_I^0=\xi^{0 \ast} \chi_I$ 
and $\phi_I^1=-\frac{1}{\sqrt{3}}( \tau \cdot \xi^{1 \ast}) \chi_I$ 
where $\xi^{0 \ast}$ is the diquark isospin-0 state 
and  $\xi^{1 m \ast}$ are the cartesian components
of the isospin-1 state with projections $m=0,\pm 1$, and $\chi_I$ is
the nucleon isospin state with nucleon isospin projection $I=\pm$1/2. 
%\be
%n=\chi^{-1/2}= 
%\left( 
%\begin{array}{c}
%0 \\1 \\
%\end{array} 
%\right) \\
%\hspace{0.5cm}
%p=\chi^{+1/2}= 
%\left(
%\begin{array}{c}
%1 \\ 0 \\
%\end{array}
%\right)
%\ee 
As explained in Refs.\ \cite{Nucleon,Gross08b} $ (\tau \cdot \xi^{1 \ast}) \chi_I$ 
generates the 3-quark isospin state in terms of the 
nucleon isospin.  

On the Dirac space, the the spin-0 component  is simply $u_{_N}(P,\lambda_n)$ (denoted simply by $u(P,\lambda_n)$ in our previous work), where $\lambda_n$ is the projection of the nucleon spin along the $z$ axis.  The spin-1 component is a vector product of the diquark polarization vectors 
$ \varepsilon_{\lambda P}^{\alpha \ast}$  and the Dirac operator  
\be
U_\alpha(P,\lambda_n)=
\frac{1}{\sqrt{3}} \gamma_5 \left(
\gamma_\alpha - \frac{P_\alpha}{m_H} \right) u_{_N}(P,\lambda_n) \label{eq:2.2}
\ee
with (generalizing the definition of $u_{_N}$ to $u_{_H}$ for an arbitrary hadron with mass $m_H$)
\be
u_{_H}(P,\lambda_H)=\sqrt{\frac{E_{_H}({\rm P})+m_H}{2m_H}}\left[ \begin{array}{c} 1 \cr \cr {\displaystyle\frac {2\lambda_h{\rm P}}{E_{_H}({\rm P})+m_H}}\end{array}\right] \chi_{_{\lambda_H}}\, , \label{eq:2.3}
\ee
where $E_{_H}({\rm P})=\sqrt{m_H^2+{\rm P}^2}$.
Note that $u_{_H}(P,\lambda_H)$ is independent of $m_H$ when ${\rm P}=0$.  

Note also that $U_\alpha(P,\lambda_n)$ satisfies the two auxiliary conditions
\bea
&&(\not\!P-m)U_\alpha(P,\lambda_n)=0
\nonumber\\
&&\qquad P^\alpha\, U_\alpha(P,\lambda_n)=0 \, .\label{eqaux1}
\eea

The S-wave component of the $\Delta$ wave function must be 
symmetric in spin and isospin. 
Because the total isospin of the $\Delta$ is 3/2  the diquark spin 0-isospin 0
component  cannot contribute, and the $\Delta$ wave function  can contain only a diquark spin-1 isospin-1 component.  As defined in Ref.~\cite{NDelta} the S-wave component of the $\Delta$ wave function is written
\be
\Psi_{\Delta}^{S }(P,k)= - \psi_{\Delta}^S(P,k) 
\tilde \phi_I^{1} \varepsilon_{\lambda P}^{\beta \ast} w_\beta(P,\lambda_\Delta).
\label{eqDelS}
\ee
In this expression $w_\beta(P,\lambda_\Delta)$ is the Rarita-Schwinger vector-spinor 
\cite{Rarita41,Milford55}  satisfying the usual auxiliary conditions
\ba
P^\beta w_\beta =0,
\nonumber\\
\gamma^\beta w_\beta=0\, ,
\label{eqaux3}
\ea
and the Dirac equation
\be
(\not\! P-M)\,w_\beta(P,\lambda_\Delta)=0\, .
\label{wdirac}
\ee
 The function
$\tilde \phi_I^{1}= (T \cdot \xi^{1 \ast}) \tilde \chi_I$ 
is the isospin part of the state, with $T$ the isospin transition operator, and $\tilde \chi_I$  the isospin-3/2 state of projection $I$.
%\ba
%\tilde \chi^{+3/2}=
%\left(
%\begin{array}{c}
%$1 \\ 0 \\ 0\\ 0 \\
%$\end{array}
%$\right)
%
%\hspace{.5cm} 
%\tilde \chi^{+1/2}=
%\left(
%\begin{array}{c}
%0 \\ 1 \\ 0\\ 0 \\
%\end{array}
%\right) \nonumber \\
%\tilde \chi^{-1/2}=
%\left(
%\begin{array}{c}
%0 \\ 0 \\ 1\\ 0 \\
%\end{array}
%\right) 
%\hspace{.5cm} 
%\tilde \chi^{-3/2}=
%\left(
%\begin{array}{c}
%0 \\ 0 \\ 0\\ 1 \\
%\end{array}
%\right) 
%\ea 

For a particle of mass $m_H$ and three-momentum 
P in the $z$ direction, the fixed-axis basis states 
used in (\ref{eqPsiN}) and (\ref{eqDelS}) are defined as
\ba
& &
\varepsilon_{\pm P}^\mu=
\mp \frac{1}{\sqrt{2}} (0,1, \pm i,0) \nonumber\\
& &
\varepsilon_{0 P}^\mu =
\frac{1}{m_H}
(\mbox{P},0,0,\sqrt{m_H^2+ \mbox{P}^2}).
\label{eq:epsilon}
\ea
These  are the diquark fixed-axis polarization states  discussed in great detail in \cite{Gross08b}. Here it is sufficient to note that
they satisfy the orthogonality condition $P \cdot \varepsilon_{\lambda P}= 0$, and that
\bea
\sum_\lambda \varepsilon^\mu_{\lambda\,P}
(\varepsilon^\nu_{\lambda\,P})^*=-g^{\mu\nu}+ \frac{P^\mu P^\nu}{m_H^2}\, .
\eea
Because of the orthogonality condition, the wave functions  
(\ref{eqPsiN}) and (\ref{eqDelS})
satisfy the  Dirac equation for the mass $m_H$.

As discussed in \cite{Gross08b} the fixed-axis 
diquark polarization states introduce 
no angular dependence in the wave function,
and therefore are convenient to describe S-states, without introducing any extra constraint.
We will show subsequently here that they are also 
convenient to define higher angular momentum states.

Finally, for the scalar wave functions $\psi_N^S$ and $\psi_{\Delta}^S$ 
in (\ref{eqPsiN}) and (\ref{eqDelS}),  which
describe  the relative quark-diquark radial motion,
we use the parametrizations
\ba
& &
\psi_N^S=\frac{N_0}{m_s(\beta_1+\chi_N)(\beta_2+\chi_N)} 
\label{eqS1}\\
& &
\psi_{\Delta}^S=\frac{N_S}{m_s(\alpha_1+\chi_\Delta)(\alpha_2+\chi_\Delta)^2},
\ea
where 
\be 
\chi_H= \frac{(P-k)^2- (m_H - m_s)^2}{m_H m_s}
\ee
for $H=N,\Delta$.
This parametrization allows for an interplay between  two different momentum
scales in the problem, quantified by the  $\beta_1$ and $\beta_2$ 
parameters for the nucleon,
and $\alpha_1$ and $\alpha_2$ for the $\Delta$.

The factors $N_0$ and $N_S$ are normalization 
constants fixed by the condition
\be
\int_k \left[ \psi_H^S(\overline{ P},k) \right]^2 =1 , 
\ee
with $\overline{ P}=(m_H,0,0,0)$ and the covariant integral $\int_k$ is defined in Eq.~(\ref{eq:intk}) below.   The normalization condition is consistent with the charge constraint 
\be
Q_I=
3 \sum_\lambda
\int_k \bar \Psi_H^S \left(\overline{ P},k\right)  j_1(0) \gamma^0
\Psi_{H}^S (\overline{ P},k), \label{eq:charge}
\ee
where $j_1(0)$ is the isospin part of the quark 
charge operator, and $Q_I$ is the charge of 
the state with isospin projection $I$ 
for either the nucleon ($H=N$ with isospin  1/2) 
or the $\Delta$  ($H=\Delta$ with isospin 3/2). 
See Refs.\  \cite{Nucleon,NDelta}  for more details.

In this paper the $\Delta$, with total angular momentum $3/2$, is composed of two positive parity subsystems (the spin 1/2 quark and the spin 1 diquark).  We will refer to the total spin ${\cal S}$ of the state as the ``core'' spin, in order to distinguish if from the total angular momentum of the state (also called the ``spin'' of the particle).   If the orbital angular momentum between the quark and diquark is zero, then the ``core'' spin of the $\Delta$ must be 3/2.  However, a state of positive parity and total angular momentum 3/2 can also be constructed if the orbital angular momentum of the constituents is a  D-wave ($L=2$) and the core spin is either ${\cal S}=1/2$ or ${\cal S}=3/2$. (For the nucleon, in contrast, the $L=2$ orbital state
 can couple only to the core spin ${\cal S}=3/2$.) The next subsections will define these two possible $\Delta$ D-states.

\subsection{The two different spin-core D-wave components of the $\Delta$ wave function}
\subsubsection{D-wave operator}

Turning to the construction of the momentum-space part of the  D-wave components of the wave function of the $\Delta$, we start by
noting that,  in relativity, inner products of three vectors (and, consequently, magnitudes of  angles)
are not Poincar\'e invariant. Hence, the operator 
${\cal D}^{\alpha \beta}$  that generates
a D-wave in the relative momentum variable $k$, only has the pure D-wave structure in the baryon rest frame.  In any other (moving) frame the intrinsic D-state will generate components in other partial
waves. Therefore, we start by defining that operator 
in the rest frame of the $\Delta$. To find its form in that frame we exploit  the two features that define a D-wave: (i) ${\cal D}^{\alpha \beta}$   is bilinear in the 
3-momentum vector part $\bf k$ in the $\Delta$ rest frame, and (ii)  the integral of ${\cal D}^{\alpha \beta}$ 
over all the possible directions of
$\bf k$ has to vanish in the rest frame.

It is convenient to introduce a four-vector that reduces to the 3-momentum $\bf k$  in the $\Delta$ rest frame.  Defining this vector for an arbitrary hadron 
\be
\tilde k^\alpha= 
k^\alpha - \frac{P \cdot k}{m_H^2} P^\alpha,
\label{eqKtil}
\ee 
where, in the hadron rest frame, $\tilde k=(0,{\bf k})$ and $\tilde k^2= -{\bf k}^2 $.  In terms of this vector, the two defining properties of $\cal D$ lead immediately to the operator
\be
{\cal D}^{\alpha \beta}(P,k)=
\tilde k^\alpha \tilde k^\beta - \frac{\tilde k^2}{3}
\tilde g^{\alpha \beta},
\label{eqDdef}
\ee
where 
\be
\tilde g^{\alpha \beta}=
g^{\alpha \beta} - 
\frac{P^\alpha P^\beta}{m_H^2}.
\label{eqclosure}
\ee
(In this discussion we suppress the subscript $H$ on both ${\cal D}$ and $\tilde k$, relying on the reader to infer the correct operator from the context.)  Note the constraint conditions
\bea
P_\alpha {\cal D}^{\alpha\beta}=0={\cal D}^{\alpha\beta}P_\beta\, .
\eea

It is convenient to work with the spherical components of ${\cal D}$, defined to be
\bea
D_{\lambda,\lambda'}=\varepsilon^{\alpha\,*}_{\lambda\,P} {\cal D}_{\alpha\beta}(P,k)\varepsilon^\beta_{\lambda'\,P}\, .  
\eea
Using the definition (\ref{eq:epsilon}) of the fixed-axis polarization states,
it is easy to see that $D$ is a hermitian matrix. 
In the hadron rest frame the matrix elements of $D$ are related directly to the
spherical harmonics $Y^L_{m_l}({\bf k})$ with $L=2$. 
A representation convenient for later applications is
\begin{align}
&
D_{\lambda,\lambda'}= \frac{\sqrt{ 8 \pi}}{3} \; {\bf k}^2  Y^2_ {m_\ell} (\hat{\bf k})\left<1\,\lambda\;2\, m_\ell\,|\,1\lambda'\,\right>\, ,   \label{sphehar1}
\end{align}
where the vector coupling (or Clebsch-Gordon -- GC) coefficients are denoted
\bea
\left<j_1\,\mu_1\;j_2\, \mu_2| j_{12}\,\mu_1+\mu_2\right>=C(j_1j_2j_{12}; \mu_1 \mu_2)\, .
\eea
Equation (\ref{sphehar1}) shows how the operator $D$ can be interpreted as the projection of the incoming direct product state of orbital angular momentum $L=2$  $\otimes$ a spin-1 vector $\varepsilon_\lambda^\alpha$, onto an outgoing vector state $\varepsilon^*_{\lambda'}$
\begin{align}
%&
&\sum_{\lambda\lambda'} \varepsilon^\alpha_{\lambda}\, D_{\lambda,\lambda'}\varepsilon^*_{\lambda'}
\nonumber\\
&\quad= \frac{\sqrt{ 8 \pi}}{3} \; {\bf k}^2 
\sum_{\lambda\lambda'} \varepsilon^*_{\lambda'}\,\left<1\,\lambda\;2\, m_\ell\,|\,1\lambda'\,\right>
Y^2_ {m_\ell} (\hat{\bf k}) \,\varepsilon^\alpha_{\lambda}\, .   \label{sphehar2}
\end{align}

\subsubsection{Spin Projection operators}

To prepare for that construction of the D-wave components of the wave function, we recall the definitions of the  spin-projection operators
${\cal P}_{1/2}$ and ${\cal P}_{3/2}$ previously used in 
Ref.~\cite{NDelta} (and in other works).  These are constructed from the operator (\ref{eqclosure}) and the operator
\be
\tilde \gamma^\alpha =
\gamma^\alpha - \frac{\not\! P\,P^\alpha }{m_H^2}\, .
\ee
The operator has the property 
\bea
\tilde \gamma^\alpha \tilde\gamma_\alpha=3\, .
\eea
In terms of these operators, the projection operators can be written 
\ba
({\cal P}_{1/2})^{\alpha \beta} &=&
\frac{1}{3} \tilde \gamma^\alpha \tilde \gamma^\beta 
%\label{eqP1/2}
\nonumber\\
({\cal P}_{3/2})^{\alpha \beta}&=&
\tilde g^{\alpha \beta} - ( {\cal P}_{1/2} )^{\alpha \beta}.
 \label{eqP32}
\ea 
For details see Refs.~\cite{NDelta,Benmerrouche89,Haberzettl98}.
We note that  ${\cal P}_{3/2}$ can be cast into the form usually found  in the literature, 
\ba
({\cal P}_{3/2})^{\alpha \beta} &=&  g^{\alpha \beta} 
-\frac{1}{3} \gamma^\alpha \gamma^\beta  
\nonumber \\
&-&
\frac{1}{3 m_H^2} 
\left( 
\not \! P \gamma^\alpha P^\beta 
+ P^\alpha \gamma^\beta \not \! P
\right),  \label{eqP32P}
\ea 
and that these spin projectors satisfy the closure and orthogonality relations
\bea
&&({\cal P}_{1/2})^{\alpha \beta}+ 
({\cal P}_{3/2})^{\alpha \beta} = \tilde g^{\alpha \beta}
\nonumber\\
&&({\cal P}_{1/2})^{\alpha \beta}({\cal P}_{3/2})_{\beta\gamma} =0=({\cal P}_{3/2})^{\alpha \beta}({\cal P}_{1/2})_{\beta\gamma}.\qquad
\label{eqProj}
\eea
Denoting the operators (\ref{eqP32}) generically by ${\cal P}_S$, one easily sees that 
\bea
&&P_{\alpha}({\cal P}_{S})^{\alpha \beta}=0
\label{eqPorth}\\
&&\left [ \not \! P,{\cal P}_{S} \right ]=0 \\
&&\gamma_\alpha\,{\cal P}_{3/2}^{\alpha\beta}=0={\cal P}_{3/2}^{\alpha\beta}\,\gamma_\beta\, .
\label{pup13}
\eea
Note also that  the state functions previously introduced in Eqs.~(\ref{eqPsiN}) and (\ref{eqDelS}) satisfy the expected eigenvector equations
\bea
&&( {\cal P}_{1/2} )_\alpha^{\;\; \beta} U_\beta = U_\alpha \nonumber\\ 
&&( {\cal P}_{3/2} )_\alpha^{\;\; \beta} w_\beta = w_\alpha. 
\eea

\subsubsection{Construction of the two possible D-state components of the $\Delta$}

Using the operator ${\cal D}$ and the spin projection operators ${\cal P}_S$ introduced above, we can now construct D-state wave functions for the $\Delta$.  Just as the S-state wave function is a matrix element of the $\Delta$ initial state with a final state consisting of a quark and a diquark in a relative S-state, the D-state wave functions are matrix elements of the $\Delta$ initial state with a final state consisting of a quark and diquark in a relative D-state.  
The construction is carried out in two steps.  
First a D-wave dependence is introduced by contracting the ${\cal D}$
operator with the elementary  S-wave Rarita-Schwinger
wave function $w_\alpha$, giving the state
\be
{\cal W}^\alpha_{\lambda_\Delta}(P,k)= {\cal D}^{\alpha\beta}(P,k) w_\beta(P,\lambda_\Delta). 
\ee

The resulting state ${\cal W}^\alpha$ satisfies the Dirac equation.  
Next, using $P_\alpha {\cal D}^{\alpha \beta}=0$ and the completeness relation
Eq.\ (\ref{eqProj}) we conclude that this ${\cal W}^\alpha$
is actually the sum (only) of two {\it independent\/} spin components
\bea
{\cal W}^\alpha_{\lambda_\Delta}(P,k)&=&  
g^{\alpha}_{\;\; \beta} {\cal W}^{\beta}_{\lambda_\Delta}(P,k)
\nonumber \\
&= & \Big[( {\cal P}_{1/2})^{\alpha}_{\;\; \beta} +  
({\cal P}_{3/2})^{\alpha}_{\; \; \beta} \Big] {\cal W}^\beta_{\lambda_\Delta} (P,k). \qquad
\label{eqdecompose}
\eea
This leads to the definition of two independent D-wave $\Delta$ wave functions
\begin{align}
\phi_{D\;2S}(\lambda\,\lambda_\Delta)&=-3\varepsilon_{\lambda {P}}^{\beta\; \ast}\;({\cal P}_S)_{\beta \alpha}\; {\cal W}^{\alpha}_{\lambda_\Delta}({P},k)
\nonumber\\
&=-3\varepsilon_{\lambda {P}}^{\beta \ast}\;({\cal P}_S)_{\beta \alpha}\; {\cal D}^{\alpha\gamma}(P,k)\,w_\gamma({P},\lambda_\Delta)
\label{eqD1a}\, ,
\end{align}
where the factor of $-3$ has been added for convenience, $S$=1/2 or 3/2, and $\varepsilon^*_\lambda$ describes the state of the outgoing diquark, just as in the S-state wave functions Eqs.~(\ref{eqPsiN}) and (\ref{eqDelS}).  Equation (\ref{eqD1a}) defines the spin part of the two D-state wave functions only; isospin and radial parts will be added below.  These wave functions satisfy the Dirac equation (\ref{wdirac}).

It is interesting to see how the wave functions (\ref{eqD1a}) have the correct spin structure corresponding to the two different  
$(L,{\cal S})$ coupling configurations, $\left(2,\frac{1}{2} \right)$ and  $\left(2, \frac{3}{2} \right)$, both giving total $J=3/2$.  Here we summarize the main points; details are given in Appendix \ref{apDwave}.  The first step is to introduce core spin wave functions (direct products of the   spin-1 diquark and a spin-1/2 quark) with ${\cal S}=1/2$ or 3/2.  These core wave functions, denoted generically by $V_S$, are constructed using CG coefficients
\bea
V_{S}^\alpha(P,\lambda_s)=\sum_\lambda\left<\sfrac12 \,\lambda \;1\,\lambda' |S\,\lambda_s\right>\varepsilon^\alpha_{\lambda'P}\, u_\Delta(P,\lambda), \label{eq:233}
\eea
where $u_\Delta$ was defined in Eq.~(\ref{eq:2.3}) (with $H\to\Delta$).
It is easy to see that the $V_{S}^\alpha$ satisfy the Dirac equation (\ref{wdirac}), that $P_\alpha V^\alpha_S=0$, and it can be shown that they are eigenstates of the projectors ${\cal P}_{S}$.  The fact that $V_{3/2}$ also satisfies the special spin 3/2 constraint $\gamma_\alpha V^\alpha_{3/2}=0$ is shown in Appendix \ref{apDwave}.

These wave functions are orthonormal and complete
\bea
\overline{V}_S^\alpha(P,\lambda){V}_{S \alpha} (P,\lambda') &=& 
\delta_{\lambda\lambda'}  \nonumber\\
\sum_{\lambda_s} V^\alpha_S(P,\lambda_s) \overline{V}^\beta_{S}(P,\lambda_s)
&=&({\cal P}_{S})^{\alpha\beta}\left[\frac{m_\Delta+\not\!P}{2m_\Delta}\right] \, .\qquad \label{eq:234}
\eea
Using the Dirac equation to introduce the projection operator into  Eq.~(\ref{eqD1a}), and then inserting these expansions, allows us to express the D-state wave functions in the form  
\begin{align}
\phi_{D\,2S}(\lambda)
&=-3\varepsilon_{\lambda P}^{\beta\, \ast}\sum_{\lambda_s} V_{S\,\beta}(P,\lambda_s)
\nonumber\\
&\qquad\qquad\times\Big\{ \overline{V}_{S\,\alpha}(P,\lambda_s){\cal D}^{\alpha\gamma}w_\gamma(P,\lambda_\Delta)\Big\}\qquad\qquad
\nonumber\\
&=(-1)^{S-\frac12}\sqrt{4\pi}\,{\bf k}^2\;\varepsilon_{\lambda P}^{\beta\, \ast}
\nonumber\\
&\qquad\qquad\sum_{m_\ell} 
 \braket{2\; m_l; S \lambda_s }{ \sfrac{3}{2} \lambda_\Delta}
Y^2_{ \; m_l}V_{S\,\beta}(P,\lambda_s)
\, .\label{eq:D1&3} 
\end{align}
This displays the two states as sums over either an $S=$1/2 or 3/2 core wave function $V_S$ times an
orbital angular momentum $L=2$ 
spherical harmonic function $Y^2_{m_l}$
coupled to a spin 3/2 $\Delta$ state, and is demonstrated in Appendix \ref{apDwave}.   Using (\ref{eq:D1&3}) and the normalization of the $V_S$ states, the normalization of the $\phi_{D\,2S}$ are 
\begin{align}
\frac{1}{4\pi}\int d\Omega_{\hat {\bf k}}\sum_\lambda |\phi_{D\;2S}(\lambda\,\lambda_\Delta)|^2&= {\bf k}^4\sum_{m_\ell} 
 \braket{2\; m_l; S \lambda_s }{ \sfrac{3}{2} \lambda_\Delta}\nonumber\\
 &={\bf k}^4\, .
\end{align}

Using the definitions  (\ref{eqD1a}) and adding the isospin factor and scalar wave function,
the complete $\Delta$ D-state wave functions 
are
\ba
& &
\Psi^{D1}_{\Delta}(P,k) =\phi_{D1} (\lambda)\;\tilde \phi_I^{1}\; 
\psi_\Delta^{D1}(P,k)   %\label{eqDelD1}
\nonumber\\
& &
\Psi^{D3}_{\Delta}(P,k) = \phi_{D3}(\lambda) \; \tilde \phi_I^{1} \;
\psi_\Delta^{D3}(P,k),
\label{eqDelD1}
\ea  
where the following simple forms were used for the scalar functions $\psi_\Delta^{D1}$ and $\psi_\Delta^{D3}$
\ba
\psi_\Delta^{D1}(P,k) &=&  N_{D1}  
\Big[
 \frac{1}{m_s^3(\alpha_3+\chi_\Delta)^4)}
 \nonumber\\
 &&\qquad-
\frac{\lambda_{D1}}{m_s^3(\alpha_4+\chi_\Delta)^4} \Big]
\label{eqP1} \\
\psi_\Delta^{D3}(P,k) &=&   \frac{N_{D3}}{m_s^3(\alpha_5+\chi_\Delta)^4}.
\label{eqP3}  
\ea
The D1 state has
two range parameters ($\alpha_3$ and $\alpha_4$) 
and the D3 state only one ($\alpha_5$). The three parameters are adjusted to the data.
%The inclusion of only one range 
%parameter $\alpha_i$ in each term  
%will simplify the interpretation of each D-state,
%although the D-state ranges 
%cannot be directly compared with the S-state ($\alpha_1,\alpha_2$)
%due to 
%the $\tilde k^2$ factor 
%(the scalar wave function vanishes for $\tilde k^2=0$).
%
We anticipate that our numerical results for the range parameters are
consistent with an expected longer range
(in $r$-space) for the D-states relatively to the S-state.  Note that the definition of ${\cal D}$
guarantees also that the D-state wave function will go as ${\bf k}^2$
when ${\bf k} \to 0$, as expected  \cite{Buck79}, and an additional mass factor $m_s^{-2}$ is introduced 
to compensate for the dimensions introduced by this  ${\bf k}^2$ dependence
of the ${\cal D}$ matrix (so that the product of ${\cal D}$ 
with $\psi_\Delta^{D1}$ or $\psi_\Delta^{D3}$ have no dimensions). 
The power 4 in the denominators of the previous equations 
was chosen to reproduce the 
expected pQCD behavior for large $Q^2$ 
($G_E^\ast \sim 1/Q^4$, $G_C^\ast \sim 1/Q^6$) 
\cite{Carlson}, 
and also to assure the convergence of 
the normalization integrals
%, compensating the 
%effect of the factor $\tilde k^2$ in the denominator.

Combining
Eqs.\ (\ref{eqDelS}) and (\ref{eqDelD1}),
the total $\Delta$ wave function can be written 
\be
\Psi_\Delta=N \left[ \Psi^S_{\Delta} + a \Psi^{D3}_{\Delta} + b\Psi^{D1}_{\Delta} 
\right],
\label{eqPsiDtotal}
\ee
where $a$ and $b$ are admixture coefficients.  The  D1 component with core spin 1/2 is orthogonal to both of the other components because of the orthogonality condition (\ref{eqProj}) and the  two components with core spin 3/2,  $\Psi^{D3}_{\Delta}$ and $\Psi^S_{\Delta}$, are orthogonal because the overlap integral is linear in $\int_k Y_{20}(z)=0$.  We will chose to normalize the individual states to unity, giving $N= 1/\sqrt{1+ a^2+ b^2}$ for the  
overall normalization factor.

\subsection{Normalization and orthogonality condition} \label{sec:ortho}

The individual D-wave scalar wave function will be chosen to satisfy the normalization conditions
\ba
& &
\int_k \left\{ \tilde k^4 \left[
\psi_\Delta^{D\,2S}(\bar P,k) \right]^2
\right\}=1.
\label{eqNormD3}
\ea
%where $S_1$ is the angular function of the variable
%$z=\cos \theta$ 
%$$S_1= \frac{1}{2}(1-z^2).$$
This determines the coefficients $N_{D1}$ (as a function of $\lambda_{D1}$) and $N_{D3}$.
 
The two components $\Psi^S_{\Delta}$ and $\Psi^{D3}_{\Delta}$ are orthogonal to the nucleon S-state $\Psi^{S}_N$, but the component $\Psi^{D1}_{\Delta}$ is, in general, not orthogonal to the nucleon S-state.  This happens because both wave functions have a core spin $S=$ 1/2, and even though the D1 state depends on $Y^2_{m_\ell}$, it  is impossible for both particles to sit simultaneously in their rest frame, so the angular integral always has some other angular dependence that prevents it from being exactly zero.  The orthogonality condition
\be
\sum_\lambda
\int_k {\bar{\Psi}}^{D1}_{\Delta}( \bar P_+,k) \Psi^{S}_N(\bar P_-,k)=0,
\label{eqOrtog}
\ee
where $\bar P_+$, $\bar P_-$ represent 
the baryon momenta for $Q^2=0$,
must be imposed numerically, and this can be done only at one value of $Q^2$.  This condition determines $\lambda_{D1}$.  As we will see below, our treatment of gauge invariance requires that we  impose the condition (\ref {eqOrtog}) at the point $Q^2=0$.  Working in the $\Delta$ rest frame, the momenta $\overline{P}_+$ and $\overline{P}_-$ 
are therefore
\bea
\overline{P}_+&=&(M,0,0,0)\nonumber\\ 
\overline{P}_-&=&\left(\frac{M^2+m^2}{2M},0,0,-\frac{M^2-m^2}{2M}\right).
\eea

To determine the coefficients  $N_{D1}$ and $\lambda_{D1}$
in the D1 component,  we first fix $\alpha_3$ and $\alpha_4$.
Then  $\lambda_{D1}$ is determined by  (\ref{eqOrtog}),
and finally  the value of $N_{D1}$ fixed by the normalization condition (\ref{eqNormD3}).

\subsection{Properties of the wave functions 
under a Lorentz transformation}

The form for the  wave functions given in Eqs.\  (\ref{eqDelD1}) 
holds only for the case where the particle 
is moving along the $z$ direction
(with 4-momentum  $P=(\sqrt{m_H^2+ \mbox{P}^2},0,0,\mbox{P})$.
The generic wave function can be obtained from an
arbitrary Lorentz transformation $\Lambda$:
\be
P^{\prime \mu}= \Lambda^{\mu}_{\; \nu} P^\nu.
\ee
%
%From the transformation laws for all of the components

Under a Lorentz transformation we obtain:
\ba
& &\varepsilon_{P^\prime}^\mu= \Lambda^\mu_{\; \nu} \varepsilon_P^\nu \nonumber\\
& &w^\prime_\beta(P^\prime)= \Lambda_\beta{}^{\alpha}S(\Lambda) w_\alpha(P)
\label{eqWtrans}
 \nonumber\\
& &u^\prime (P^\prime) = S(\Lambda) u(P) 
\label{eqUtrans}
\nonumber\\
& & {\cal D}_{\alpha \beta} (P^\prime,k^\prime)=
\Lambda_\alpha^{\; \sigma} 
\Lambda_\beta^{\; \rho} {\cal D}_{\sigma \rho} (P,k) \nonumber\\
& & 
S^{-1}(\Lambda) ({\cal P}_S^\prime)_{\alpha \beta} 
S(\Lambda) =
\Lambda_\alpha^{\; \sigma} 
\Lambda_\beta^{\; \rho} 
({\cal P}_S)_{\sigma \rho},
\label{eqBigWtrans}
\ea
where $u^\prime$ and $w_\beta^\prime$ 
represents the states in the arbitrary frame.
For simplicity,
the dependence of the spinor states on the Wigner rotations acting on the polarization vectors has not been shown explicitly, and $({\cal P}_S)$ are the projectors of (\ref{eqP32}) with  $({\cal P}_S^\prime)$ the same projectors with $P^\prime= \Lambda P$, one obtains the transformation law
\ba
{\cal Z}^\prime_{\beta}(P^\prime,k^\prime)=
S(\Lambda) \Lambda_\beta^{\; \alpha} {\cal Z}_\alpha(P,k)
\label{wtrans}
\ea
for any vector spinor state ${\cal Z}$.  Finally, from (\ref{wtrans}) the transformation laws for the  total $\Delta$ wave function follows
\be
\Psi_\Delta^\prime (P^\prime,k^\prime)=
S(\Lambda) \Psi_\Delta(P,k).
\ee

In conclusion, we may derive the baryon  wave function in any frame, 
where the four-momentum $P$ is arbitrary, 
by means of a Lorentz transformation $\Lambda$ on the wave function
defined in the baryon rest frame.

\section{Form factors for the $\gamma N \to \Delta$ transition }
\label{secFFgen}
\subsection{Definitions}

The electromagnetic $N \Delta$ transition current  is
%(the electron charge $e$) is given by 
\be
J^\mu = \bar w_\beta(P_+) \Gamma^{\beta \mu} 
(P,q)\gamma_5 u(P_-)\,\delta_{_{I'I}},
\label{eqJ1}
\ee
where $P_+$ ($P_-$) is the momentum of the $\Delta$ (nucleon), $I'$ ($I$) the isospin projection of the $\Delta$ (nucleon), and 
the operator $\Gamma^{\beta \nu}$ can 
be written in general \cite{Jones73} as 
\be
\Gamma^{\beta \mu } (P,q) = G_1 q^\beta \gamma^\mu +G_2 q^\beta P^\mu +
G_3 q^\beta q^\mu- G_4 g^{\beta \mu}.
\label{eqJS}
\ee 
Although we have omitted the helicity indices for these
states, the transition current depends on both the helicities of the final and initial baryons and
on the photon helicity.
The variables $P$ and $q$ are respectively the average of 
baryon momenta and the absorbed (photon) momentum: 
\ba
& &P=\sfrac{1}{2}\left({P_++P_-} \right) %\label{eqP}
\nonumber\\
& &q=P_+-P_-. \label{eqQ}
\ea

The form factors $G_i$, $i=1,..,4$ are functions of 
$Q^2=-q^2$ exclusively.
Because of  current conservation, $q_\mu \Gamma^{\beta \mu} =0$, only 
three of the four form factors are independent. 
In particular, we can write $G_4$ in terms of the other 
three form factors as 
\be
G_4=(M+m) G_1 +\frac{M^2-m^2}{2}G_2 -Q^2 G_3,
\label{eqG4}
\ee
and adopt the structure originally proposed by 
Jones and Scadron \cite{Jones73}.
%[Equation (\ref{eqG4}) holds between asymptotic states.] 
Alternatively (see below) we can write $G_3$ in terms of the other three
\be
G_3=\frac{1}{Q^2}\Big[(M+m) G_1 +\frac{M^2-m^2}{2}G_2 - G_4\Big].
\label{eqG3}
\ee

The parametrization (\ref{eqJS}) 
in terms of the form factors $G_i$ is not 
the most convenient one for comparison with the experimental data.
More convenient are the magnetic dipole (M), 
electric quadrupole (E) and Coulomb quadrupole (C) 
form factors.  These can be defined directly in terms of helicity amplitudes 
 \cite{Carlson,Jones73}.
{\it Note that the form factor $G_3$ does not enter directly into the expressions for the helicity amplitudes\/} because $\epsilon_\lambda^{\mu\,*}q_\mu=0$ for all $\lambda$.  But, if we use the constraint (\ref{eqG4}) to eliminate $G_4$, $G_3$ appears in these expressions and we obtain 
\ba
G_M^\ast(Q^2)&=& \kappa
\left\{ \frac{}{}
\left[ (3M +m)(M+m) +Q^2 \right] \frac{G_1}{M} \frac{}{}
\right.\nonumber  \\
& +&
\left. 
(M^2-m^2) G_2 -2 Q^2 G_3  \frac{}{}
\right\}   \label{eqGM} \\
G_E^\ast(Q^2)&=& \kappa
\left\{ (M^2 -m^2 -Q^2) \frac{G_1}{M} \frac{}{}
\right.  \nonumber \\
& +& 
\left.  
(M^2-m^2) G_2 -2 Q^2 G_3  \frac{}{}
\right\}  \label{eqGE} \\
G_C^\ast(Q^2)&=& \kappa
\left\{ 4M G_1 +(3M^2+m^2+Q^2)G_2   \frac{}{}
\right.  \nonumber \\
&+& 
\left. 
2 (M^2-m^2-Q^2) G_3   \frac{}{}
\right\},  \label{eqGC}
\ea
where 
\be
\kappa= \frac{m}{3(M+m)}.
\ee
These three form factors $G_a^\ast$ ($a=M,E,C$) 
are, respectively, the magnetic, electric and Coulomb (or scalar) 
multipole transition form factors.
%and will be mention as the physical form factors.

As $G_M^\ast$ dominates 
at low momentum $Q^2$, the following ratios 
are useful
\be
R_{EM}(Q^2)=-\frac{G_E^\ast(Q^2)}{G_M^\ast(Q^2)},
\label{eqREM}
\ee
and
\be
R_{SM}(Q^2)=-\frac{|{\bf q}|}{2M}\frac{G_C^\ast(Q^2)}{G_M^\ast(Q^2)},
\label{eqRSM}
\ee
where ${\bf q}$ is the photon 3-momentum in the $\Delta$ rest frame
%$|{\bf q}|$ can be written as 
%\be
%|{\bf q}|=\frac{\sqrt{\left[(M+m)^2+Q^2 \right] 
%\left[(M-m)^2 +Q^2 \right]}}{2M}.
%\ee
\be
|{\bf q}|=\frac{\sqrt{d_+ d_-}}{2M},\label{eqqm}
\ee
with 
\bea
d_\pm= (M\pm m)^2+ Q^2\, . 
\label{eqdpm}
\eea

The analysis of the transition at large 
$Q^2$ in the pQCD regime 
(where quarks and gluons are the appropriate 
degrees of freedom) %where vector dominance is expected, 
gives $G_M^\ast \simeq -G_E^\ast \sim 1/Q^4$ 
and $G_C^\ast \sim 1/Q^6$ \cite{Carlson}.

\subsection{The $G_1, G_2, G_3$  set  versus the $G_1, G_2,G _4$ set}

The representation 
of  the electromagnetic current in terms of the 3
independent $(G_1,G_2,G_3)$  form factors,
as proposed by Jones and Scadron \cite{Jones73},  is not the most convenient
choice that can be made.  As mentioned above, the form factor $G_3$  is not part of the helicity transition amplitudes given  by the operator $\varepsilon_\mu(q) J^\mu$, due to the 
condition $\varepsilon  \cdot q=0$. 
For this reason it seems natural to 
replace the set $(G_1,G_2,G_3)$ by $(G_1,G_2,G_4)$.
In this basis,  $G_M^\ast$, $G_E^\ast$, and $G^*_C$ are are given by 
\ba
G_M^\ast&=&\kappa 
\Big\{ 2 G_4+d_+\,\frac{G_1}{M} \Big\}
\label{eqGM1}\\\
G_E^\ast&=& \kappa 
\Big\{ 2 G_4 - d_+\,\frac{G_1}{M} \Big\} 
\label{eqGE1}\\
G_C^\ast &=&
\frac{\kappa}{Q^2} 
\Big\{
2(M-m)\,d_+ G_1 +  
d_+ d_- G_2  \nonumber\\
& &\qquad
-2(M^2-m^2-Q^2) G_4 \Big\} 
\label{eqGEGM}
\ea
where $d_\pm$ were defined in Eq.~(\ref{eqdpm}).  
Note that the multipole form factors 
$G_M^\ast$ and $G_E^\ast$ do not depend on $G_2$.

Eq.~(\ref{eqGEGM}) for $G_C^*$
presents an apparent singularity when $Q^2 =0$.  The presence of this apparent
singularity is the historical reason for  choosing $G_1$, $G_2$, $G_3$ to be
the independent form factors; this choice gives finite form factors under any circumstances.
  However, if the theory conserves current, {\it with a $G_3$ that is finite at $Q^2=0$\/} (a required feature of any consistent model), then the singularity disappears as $Q^2 \to 0$, since, using the current conservation condition (\ref{eqG3}), the numerator (at $Q^2=0$) is proportional to 
\ba
\left[(M+m) G_1+ \frac{M^2-m^2}{2}G_2-G_4 \right]
=  {Q^2 G_3},\qquad \label{eqconstraint}
\ea
which, if $G_3$ is finite, approaches zero as $Q^2\to0$.

We prefer the independent choice $G_1$, $G_2$, $G_4$ because it enables us to discuss the restrictions imposed by current conservation in a more 
transparent way.  
Many models do not automatically conserve current 
(this is true for our D1 component, as we will discuss below).   
If we start with a model that does 
not naturally conserve current, we prefer to impose current 
conservation by modifying the current in the following way:
\be
J^\mu \to J^\mu + \frac{(q \cdot J)}{Q^2}q^\mu.
\label{eqcurrentcon}
\ee
This way of imposing current conservation is, of course, not unique, but has the nice property that the additional term added is proportional to $q^\mu$, and hence makes {\it no additional contributions to any observables obtained by contracting the current with another conserved current or with a photon polarization vector\/}, always orthogonal to $q^\mu$ (in the Lorentz gauge, our choice).  In the application discussed in this paper, the modification (\ref{eqcurrentcon}) will only alter the $G_3$ form factor, and when we use the expressions (\ref{eqGM1}) -- (\ref{eqGEGM}) we see that they are unchanged by any modification of $G_3$.  Hence, our method allows us to choose $G_3$ to satisfy current conservation, without changing the basic predictions of the theory.

However, current conservation is like the Cheshire cat, while
the consequences of imposing it seem to have vanished, a ``smile'' still remains. 
What remains is the requirement that there is
no singularity in $G_C^*$ as $Q^2\to0$.
This requirement is satisfied by modifying the form factors in such a way that the linear combination (\ref{eqconstraint}) is zero at $Q^2=0$.  Implementation of this requirement will be discussed below.

\subsection{Simple relation for  $G_C^\ast$  }

In the following discussion we will work in the rest frame of the outgoing $\Delta$, where the four-momenta (\ref{eqQ}) become
\ba
& &
q^\mu=(\omega,0,0,|{\bf q}|) \nonumber\\
& &
P^\mu=\left(\frac{2M-\omega}{2},0,0,-\frac{|{\bf q}|}{2} 
\right),
\ea
where $|{\bf q}|$ was given in Eq.~(\ref{eqqm}), and the photon energy $\omega$ can be written 
in terms of the nucleon %on-mass-shell 
energy 
$\omega=M-\sqrt{m^2+|{\bf q}|^2}$, or
\be
\omega= \frac{P_+ \cdot q}{M}=
\frac{M^2-m^2-Q^2}{2M}.
\ee
In this frame the photon moves in the $+\hat z$ direction, with polarization vectors
\ba
& &
\epsilon_{\pm\, q}^\mu=
\mp \frac{1}{\sqrt{2}} (0,1, \pm i,0) \nonumber\\
& &
\epsilon_{0 \,q}^\mu =
\frac{1}{Q}
(|{\bf q}|,0,0,\omega).\label{eqphoton}
\ea
Note that the transverse states ($\lambda=\pm1$) are identical to those defined in Eq.~(\ref{eq:epsilon}), but that the longitudinal state is very different.  All of these satisfy the constraint $q_\mu \epsilon^\mu_\lambda=0$, and because $q^z>0$ are identical to helicity states.  While we will work out the explicit relations in this rest frame, all relations that are derived from four-vector scalar products are, of course, independent of the frame.

Introduce the photon helicity amplitudes of the electromagnetic transition current (\ref{eqJ1}) (for a general discussion of helicity amplitudes see Refs.\ \cite{Carlson,Jones73})
\bea
\epsilon^\mu_{\lambda\,q}\,J_\mu={\cal J}_\lambda
\eea
where the polarizations of the $N$ and $\Delta$ will remain unspecified.  Note immediately that $G_3$ does not contribute to any of these amplitudes, and because $P\cdot\epsilon_\pm=0$, the transverse amplitudes do not depend on $G_2$.  The only amplitude that depends on $G_2$ is the longitudinal ${\cal J}_0$.  Using the relations
\bea
&&\epsilon_{0\,q}^\mu P_\mu=\frac{|{\bf q}|M}{Q}\equiv \frac{1}{a_P}
\nonumber\\
&&\epsilon_{0\,q}^\mu=a_q q^\mu + a_PP^\mu;\quad a_q=\frac{M^2-m^2}{2|{\bf q}|QM}\qquad\label{eqconditions}
\eea
we can reduce the terms $\not\!\!\epsilon_{0\,q}$ and $\epsilon^\beta_{0\,q}$ that occur when using (\ref{eqJ1}) to evaluate ${\cal J}_0$, and obtain
\be
{\cal J}_{0\;s^\prime s}= {\cal R}_{s^{\prime} s}\frac{(M+m)}{2m} \frac{3\,Q}{\sqrt{d_+d_-}}\,G_C^*
\label{eqDel0}
\ee
where $s$ and $s^\prime$ are, respectively, 
the nucleon and $\Delta$ spin projections along the $z$-axis, and  
\ba
{\cal R}_{s^{\prime} s} &=& \bar w_\beta(P_+,s') q^\beta \gamma_5 u(P_-,s)
 \nonumber \\
        &=&\delta_{s s^\prime}(2s) \sqrt{\frac{2\,d_+}{3mM}} \frac{d_-}{4M} .\qquad
\ea
We emphasize that, provided we use Eq.~(\ref{eqGEGM}) to define $G_C^*$, this relation holds for all models,  {\it even those that do not conserve current\/}.  Note that ${\cal R} \ne 0$ only if the spin projections are equal 
($s=s^\prime$).  We may conclude that $G_C^\ast\ne0$ only if (for example) ${\cal J}_{0\;\frac{1}{2} \frac{1}{2}}\ne0$, and using Eq.\ (\ref{eqDel0}) we obtain
\bea
{\cal J}_{0\;\frac12 \frac12}=\frac{(M+m)}{4m}\sqrt{\frac{3\,d_-}{2\,mM}}\left[\frac{Q}{M}\right]G^*_C \, .\label{eqGCJ0}
\eea

\section{The electromagnetic current within the Spectator Model}
\label{secEMcur}

In this section we study  how the electromagnetic current can be
constructed within the constituent quark model (CQM) for the baryon structure
presented in Sec.~\ref{secWave Function}. 

In any CQM  model the quarks making up the baryons are 
not point particles, but 
composite valence quarks, dressed by their gluon and 
sea quark structure.  Here we use the Covariant Spectator Theory and assume  the baryon is  a quark-diquark system,
as explained in Sec.~\ref{secWave Function}.

The on-shell diquark mass $m_s$ 
scales out from the
elastic form factor, which turns out to be 
independent of the diquark mass \cite{Nucleon}.   This mass does not scale out of the deep inelastic results (DIS) and the qualitative description of DIS 
leads to the estimate of $m_s\simeq 0.8m$, 
allowing a natural interplay between low and 
high energy phenomenology. 
This interplay is needed since the factorization into
low and high energy scales does not apply exactly.

In the following we will explain how 
gauge invariance conveniently constrains the
current, when the internal structure 
of the quarks is parametrized in terms 
of phenomenologically fixed wave functions.

\subsection{Implications of the choice of current}

\subsubsection{Simple current}

Constituent quarks are dressed particles with a complex effective structure,
an effective charge and magnetic moment. Therefore, their current
consists of a Dirac and a Pauli  term, and can be written as
\be
j_{I\,a}^\mu=
j_1 \gamma^\mu 
+ j_2 \frac{i \sigma^{\mu \nu}q_\nu}{2m}.
\label{eqJI}
\ee
(The subscript ``$a$'' on the current will be dropped in subsequent discussion, and will be used only when we need to distinguish this current from the modified current discussed in the next subsection.)
The form factors $j_1$ and $j_2$ 
are normalized in order to describe the 
nucleon charge and magnetic moments 
(as functions of the quark isospin $I$) as discussed
in  Ref.~\cite {Nucleon}. 
The explicit formulas are defined by the 
Eqs.~(\ref{eqJi}) and (\ref{eqf1m}) below.
%Their explicit formulas  were presented in that
%reference.
%
The quark current (\ref{eqJI}) is not of the most general 
form. In the next subsection we will consider a more generic case,
in light of the discussion on gauge invariance 
that unfolds immediately here as consequence of (\ref{eqJI}).

To start this discussion, given the quark current (\ref{eqJI}) and the 
nucleon ($\Psi_N$) and $\Delta$ ($\Psi_\Delta$) 
wave functions, we  write the transition
current between these states.
With a positive parity axial diquark 
the only allowed states for the nucleon and $\Delta$ are  S and D states, since P-states are ruled out (unless they are associated with the lower relativistic components, not discussed so far in this series of papers).

To simplify the formulas  we 
will exclude the isospin from the discussion
(later in this paper we show how to 
include the isospin explicitly).
In impulse approximation 
\cite{Nucleon,NDelta,Stadler97a,Stadler97b,Gross04b,Adam97}
the transition current takes the form 
\be
J^\mu =
3 \sum_{\lambda} \int_k 
\bar \Psi_\Delta j_I^\mu \Psi_N,
\label{eqJbar}
\ee
where all momenta and spin projections ($s'$ for the $\Delta$ and $s$ for the nucleon)  have been suppressed.   The factor 3 sums up the contributions 
of the three quarks, the sum is over all intermediate polarizations $\lambda$  of the diquark,  and 
\be
\int_k \equiv  \int \frac{d^3 k}{(2\pi)^3 2 E_s}, \label{eq:intk}
\ee
is the covariant integral with 
$E_s=\sqrt{m_s^2+{\bf k}^2}$ as the diquark 
on-mass-shell energy.
The initial and final momentum dependence 
are not explicitly included for simplicity.

\begin{figure}[t]
\vspace{0.0cm}
\centerline{
\mbox{
\includegraphics[width=2.8in]{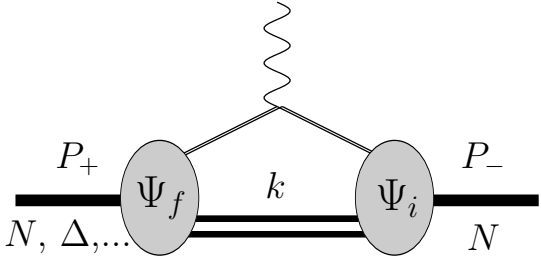}}}
\vspace{0.9cm}
\caption{\footnotesize{
Relativistic impulse approximation.}}
\label{figRIA}
\end{figure}

As discussed in \cite{Gross08b} equation (\ref{eqJbar}),  for the states 
$\Psi_\Delta$ and $\Psi_N$ defined here, 
goes beyond the scope of the relativistic impulse approximation
(RIA) shown diagrammatically in Fig.~\ref{figRIA}, 
and includes some effective two body currents.

Since both the final and initial states satisfy the 
Dirac equation: $\not\!\!P_+ \Psi_\Delta= M \Psi_\Delta$ and
$\not\!P_- \Psi_N= m \Psi_N$,
the Pauli current  
can be simplified using  the Gordon decomposition 
\ba
\sum_{\lambda} \int_k 
\bar \Psi_\Delta  \frac{i \sigma^{\mu \nu} q_\nu}{2m} \Psi_N &=&
\frac{M+m}{2m}
\sum_{\lambda} \int_k 
\bar \Psi_\Delta \gamma^\mu \Psi_N \label{eqGordon} \\
&- & 
\frac{(P_++P_-)^\mu}{2m}
\sum_{\lambda} \int_k 
 \bar \Psi_\Delta  \Psi_N. \nonumber 
\ea
The last term is proportional to (inserting the spin projections for clarity)
\be
\rho_{s's} (Q^2)\equiv\sum_{\lambda} \int_k   \bar \Psi_{\Delta\,s'}\;  \Psi_{N\,s}.
\ee
With this definition (dropping references to $s'$ and $s$ again), we can use the Gordon decomposition to write the current (\ref{eqJbar}) as 
\ba
J^\mu &=&
3 j_v
\sum_{\lambda} \int_k   \bar \Psi_\Delta  \gamma^\mu \Psi_N 
%\nonumber \\
%&- &
-3 j_2 
\frac{P^\mu}{m} \rho  (Q^2),
\label{eqJgen}
%+3 j_1 \frac{(M-m)}{Q^2} q^\mu \rho.
\ea
where 
\bea
j_v=j_1 + \frac{M+m}{2m} j_2\, . \label{eq:jv}
\eea
These equations hold in any frame.

Next, using Eq.~(\ref{eqJgen}), the relations (\ref{eqconditions}), and the Dirac equation, we find an alternative form for the longitudinal current
\begin{align}
{\cal J}_0&=
3 j_v
\sum_{\lambda} \int_k   \bar \Psi_\Delta  \not\!\epsilon_{0\,q} \Psi_N 
%\nonumber \\&
-3 j_2 
\frac{|{\bf q|}M}{Q\,m} \rho  (Q^2)
\nonumber\\
&=  \frac{3}{Q}\sqrt{\frac{d_-}{d_+}}(M+m)\,j_C\,\rho(Q^2)
\label{eqJ02}
\end{align}
where 
\bea
j_{C}=j_1 - j_2\frac{Q^2}{2m(M+m)}\, . \label{eq:jc}
\eea
Both the current ${\cal J}_0$ and the factor $\rho$ depend on the spin projections $s'$ and $s$, suppressed so far.  Taking the spin projections $s=s'=\frac12$ and combining this result with Eq.~(\ref{eqGCJ0}) gives $G_C^*$ in terms of $\rho_{_{\frac12\frac12}}$
\begin{align}
G_C^*(Q^2)&=\frac{4mM}{Q^2}\sqrt{\frac{6Mm}{d_+}}
 \,j_{C}\,\rho_{_{\frac12\frac12}}(Q^2).
\label{eqJ02aa}
\end{align}
This result holds in any frame.

Now we connect some of these results to the divergence of the 
simple current (\ref{eqJI}).  
Noting that the Pauli term, proportional to $j_2$, is automatically conserved, the divergence of current depends only on the behavior of the Dirac term, proportional to $j_1$.  Evaluating the divergence gives 
\ba
q \cdot J &=& 
3 j_1 \sum_\lambda \int_k \bar \Psi_\Delta 
\not \! q \,\Psi_N \nonumber  \\
&=& 
3 (M-m)\, j_1 \; \rho (Q^2). 
\label{eqqJ1}
\ea
Because the  masses are different ($M \ne m$) this (frame independent) result shows that the {\it simple current\/} (\ref{eqJI})  will be conserved for an electromagnetic 
transition from a state $\Psi_N$ to a state
$\Psi_\Delta$ if  and only if $\rho (Q^2)=0$.  We showed in Ref.~\cite{NDelta} that this term will vanish identically (for all values of $Q^2$) if the core spins of the two states are different.   This is true for the nucleon S-state to $\Delta$ S-state transition, and also for the transition from the nucleon S-state to the $\Delta$ D3 state.  It is not true for the transition to the D1 state, as discussed briefly above, and in more detail below.

Combining Eqs.~(\ref{eqqJ1}) and (\ref{eqJ02aa})  gives the following interesting connection 
\begin{align}
q \cdot J_{\frac12 \frac12} &= 3(M-m)\frac{ j_1}{j_{_C}}\,\sqrt{\frac{d_+}{6mM}}\,\frac{Q^2}{4mM}\, G_C^* . 
\label{}
\end{align}

The consequence of this equation is 
that the simple current (\ref{eqJI}) will be  conserved  if and only if  $G_C^\ast=0$.  Alternatively, since transitions to the 
$\Delta$ S and D3 states conserve the current (\ref{eqJI}), these {\it cannot\/} give a non-zero $G_C^\ast$.  To build a model in which $G_C^\ast\ne0$ we must find a different current, and this leads us to the next subsection.

\subsubsection{Modified current}

Following a previous work \cite{Nucleon} we replace the quark current (\ref{eqJI}) by
\be
j_I^\mu=
j_1 \left(\gamma^\mu - \frac{\not\! q q^\mu}{q^2}
\right) 
+ j_2 \frac{i \sigma^{\mu \nu}q_\nu}{2m}=j_{I\,a}^\mu + \Delta j_I^\mu\, . \label{eqJI2}
\ee
It is easy to evaluate the additional term,
\bea
\Delta j_I^\mu= 3(M-m)\frac{q^\mu}{Q^2}\,\rho(Q^2).
\label{eqcorrC}
\eea
This shows that all the {\it good\/} properties of 
the previous current remain intact; when $\rho=0$ for all $Q^2$ values,  the Dirac current $j_1\gamma^\mu$ is conserved and the correction term  vanishes identically. 
The advantage of the current (\ref{eqJI2})
is that current conservation is guaranteed, even
when $\rho(Q^2)$ does not vanish identically. 

The only possible problem with the new current is that  it might be singular at $Q^2=0$.  This singularity must be removed by imposing the requirement that  ${\rho}(Q^2) \to Q^2$ as $Q^2\to0$.   Since $\epsilon_{0\,q}\cdot q=0$, Eqs.~(\ref{eqJ02}) and (\ref{eqJ02aa}) are unchanged, and this requirement also means that $G_C^*$ is finite at $Q^2=0$, guaranteeing that the apparent singularity in Eq.~(\ref{eqGEGM}) does indeed cancel.  

The condition that guarantees that $\rho=0$ at $Q^2=0$ was already introduced above in Sec.~\ref{sec:ortho}, Eq.~(\ref{eqOrtog}).  
The importance of this orthogonality condition was emphasized in Ref.~\cite{Noble78}; imposing it ensures that the current (\ref{eqJI2}) is well defined and  conserved for all $Q^2 $.

To summarize: in the present model the orbital angular momentum 
states are not derived from an underlying hamiltonian. 
Therefore the $\Delta$ state with $(L,{\cal S})=\left(2,\frac{1}{2}\right)$, 
with the same core spin quantum numbers as the nucleon  state,
even though carrying the correct spin-isospin symmetries,  
does not have a spacial scalar part $\psi^{D1}_\Delta$ 
that is ab-initio orthogonal to a nucleon state.  
The orthogonality is imposed by  a judicious choice of
the parameter $\lambda_{D1}$ in Eq.~(\ref{eqP1}).

\subsection{Isospin dependence of the current}

For simplicity,  we did not include isospin in the discussion in the previous subsection.  It is included in the definition of the current through
the following isoscalar and isovector decomposition, as in Ref.~\cite{Nucleon}:
\ba
& &j_i= \sfrac{1}{6} f_{i+}(Q^2) + \sfrac{1}{2} f_{i-}(Q^2)\tau_3 %\nonumber\\
%& &j_2= \frac{1}{6} f_{2+}(Q^2) + \frac{1}{2} f_{2-}(Q^2)\tau_3.
\label{eqJi}
\ea
where $i=1,2$ and  $f_{1\pm}$ and $ f_{2\pm}$ were 
adjusted by the charge and magnetic form factors of the nucleon and were normalized to $f_{1\pm}(0)=1$, $f_{2\pm}(0)= \kappa_\pm$.  Only the isovector form factors, $f_{i-}$ contribute to the $\gamma N\to\Delta$ transitions. 

The overall isospin factor can be calculated separately, and was worked out in Ref.\ \cite{NDelta}.  This factor is
\bea
C_{_{I'I}}&\equiv& \tilde \phi_{I'}^{1}\frac{\tau_3}{2}\phi_{I}^{1}= -\frac{1}{\sqrt{3}}\sum_i \chi_{_{\Delta\,I'}}^\dagger T^i\frac{\tau_3}{2}\tau^i \chi_{_{N\,I}}
\nonumber\\
&=&-\frac{\sqrt{2}}{3}\delta_{_{I'I}}=C_0\;\delta_{_{I'I}}\,.\qquad
\eea
All formulas derived in the precious sections
still valid if we replace 
\ba
 & &
j_i \to C_0\; f_{i-} \, .%\nonumber \\
%& &
%j_2 \to  -\frac{\sqrt{2}}{3} f_{2-}. \nonumber 
\ea
%For details consult   Note that only the isovector quark currents contribute to the $\gamma N \to \Delta$ transition.

\section{Valence quark contribution for the Form Factors}
%\section{Form Factors in a valence quark model}
\label{secFFval}

The impulse approximation for $\gamma N \to \Delta$ 
transitions from the nucleon S-state to each of the 
$\Delta$ states can be written, 
using Eq.~(\ref{eqPsiDtotal})
\be
J^\mu= N \left[ J_S^\mu + a J_{D3}^\mu + b J_{D1}^\mu \right],
\ee
where the index identifies the $\Delta$ state.  
From this we can calculate the form factors $G_1$, $G_2$, 
and $G_4$ defined in Eqs.\  
(\ref{eqJ1}) and (\ref{eqJS}), and using the definitions  given in Eqs.~(\ref{eqGM1})-(\ref{eqGEGM}) calculate the multipole transition 
form factors $G_M^\ast, G_E^\ast$ and $G_C^\ast$.

\subsection{Transition to the $\Delta$ S-state}

The transition current from the nucleon S-state for 
the $\Delta$ S-state was already evaluated in the 
Ref.\ \cite{NDelta}.
Using the upper index $S$ to indicate the 
$\Delta$ state, the results are 
\ba
& &
G_M^S (Q^2)= \frac{8}{3\sqrt{3}} 
\frac{m}{M+m} f_v {\cal I}_S, \\
& &
G_E^S (Q^2)=0
\nonumber\\
&& G_C^S (Q^2)=0,
\ea
where $j_v$ is the analogue of (\ref{eq:jv})
\bea
f_v= f_{1-}+ \frac{M+m}{2m} f_{2-}
\eea 
and 
\be
{\cal I}_S = \int_k \psi_\Delta^S(P_+,k) \psi_N^S(P_-,k),
\ee 
is the overlap integral of the radial (scalar) wave functions. 
Asymptotically we have $G_M^S \sim 1/Q^4$, 
as showed in Ref.~\cite{NDelta}.

According to Eqs.\ (\ref{eqGE1}) and (\ref{eqGEGM}), $G_E^S$ and $G_C^S$ vanish because the terms involving $G_1$, $G_2$, and $G_4$.   
cancel exactly.

\subsection{Transitions to the $\Delta$ D-states}

The transition currents to the D states are
\bea
J_{D\,2S}^\mu &=&
3 \sum_\lambda \int_k \bar \Psi_\Delta^{D\,2S}
(P_+,k) j_I^\mu \Psi_N(P_-,k)\nonumber\\
&=& \overline{w}_\beta(P_+)\Gamma_{D\,2S}^{\beta\,\mu}(P,q)\gamma_5 u_{_N}(P_-)\,\delta_{_{I"I}}
\eea
where we suppress all reference to the spins of the nucleon and 
$\Delta$.  Substituting for $\Psi_N$ using Eq.~(\ref{eqPsiN}), $\Psi_\Delta^{D\,2S}$ using Eqs.~(\ref{eqD1a}) and (\ref{eqDelD1}), and using the general reduction (\ref{eqJgen}) gives
\begin{align}
\Gamma_{D\,2S}^{\beta\,\mu}(P,q)&=
-3\sqrt{\frac{3}{2}}C_0
\int_k \Bigg\{
{\cal D}^{\beta\beta'}(P_+,k) ({\cal P_S})_{\beta'\alpha'}\,\sum_{i=1}^2{\cal O}_i^\mu
\nonumber\\
&\quad\times \Delta^{\alpha'\alpha}
\left(\gamma_\alpha +\frac{(P_-)_\alpha}{m}\right)\Bigg\} 
\,\psi_\Delta^{D\,2S}\psi_N^S,
\end{align}
where $\Delta^{\beta\alpha}$ is the sum over the fixed axis diquark polarizations (previously derived in Ref.\ \cite{Gross08b,NDelta})
\begin{align}
\Delta^{\alpha' \alpha} =& \sum_\lambda 
\varepsilon_{\lambda P_+}^{\alpha'}  \varepsilon_{\lambda P_-}^{\alpha \ast}
\nonumber\\
=&
-\left(g^{\alpha'  \alpha} - \frac{P_-^{\alpha'}  P_+^\alpha}{b} \right)  + 
\nonumber\\
&a \left[ P_-^{\alpha'}  -\frac{b}{M^2}P_+^{\alpha'}   \right]
 \left[P_+^\alpha -\frac{b}{m^2}P_-^\alpha  \right],
\end{align}
with 
\bea
a&=& -\frac{Mm}{b(Mm+b)}\nonumber\\
b&=&P_+ \cdot P_-\, , 
\eea
and the two current operators emerging from the reduction (\ref{eqJgen}) are
\bea
{\cal O}^\mu_1&=&f_v\;\gamma^\mu
\nonumber\\
{\cal O}^\mu_2&=&-f_{2-}\frac{P^\mu}{m}\, .
\eea
Using the conditions (\ref{eqaux1}),  (\ref{eqaux3}), and (\ref{eqPorth}), the part of the expression for $\Gamma_{D\,2S}$ in curly brackets $\{\;\}$ reduces to
\begin{align}
\Gamma_{D\,2S}^{\beta\,\mu}(P,q)&=
-\sqrt{3} \int_k\Bigg\{
{\cal D}^{\beta\beta'}(P_+,k) ({\cal P_S})_{\beta'\alpha}\,\sum_{i=1}^2{\cal O}_i^\mu
\nonumber\\
&\qquad\times
\left(\gamma^{\alpha} -\frac{P_-^{\alpha}[\not\!P_+-M]}{mM+b}\right)\Bigg\} \, 
\psi_\Delta^{D\,2S}\psi_N^S. 
\label{eq:511}
\end{align}
This general expression may be reduced further by noting that, in a collinear frame in which none of the momenta have components in the $\hat x$ or $\hat y$ directions, the only dependence of the integrand on the azimuthal angle $\varphi$ is in the angular dependent term ${\cal D}$.  Hence, we may average over this angle using the (covariant) identity
\be
\frac{1}{2\pi}\int d\varphi \;{\cal D}^{\alpha \beta}(P_+,k)=  
b(\tilde k, \tilde q)\bar R^{\alpha \beta},
%\label{eqIntPHI}
\ee
where 
\ba
& &
b(\tilde k, \tilde q) =
\frac{3}{2} \frac{(\tilde k \cdot \tilde q)^2}
{\tilde q^2}
-\frac{1}{2} \tilde k^2 
%\label{eqBtilde}
\nonumber\\
& &R^{\alpha \beta} (P_+,P_-)
=
\frac{\tilde q^\alpha  
\tilde q^\beta}{\tilde q^{2}}
-\frac{1}{3} \tilde g^{\alpha \beta}
%\label{eqRtilde}
\ea
with $\tilde k$ and $\tilde q$ defined as in Eq.~(\ref{eqKtil}) [with the substitutions $P\to P_+$ and $m_H\to M$].  This identity is proved in Appendix \ref{apIntK}.

Using the conditions  (\ref{eqaux3}) and (\ref{eqPorth}) again, the $\varphi$ average of (\ref{eq:511})  can be simplified
\begin{align}
\overline{\Gamma}_{D\,2S}^{\beta\,\mu}(P,q)&=
-\sqrt{3}\;  \int_k 
\left\{
b(\tilde k, \tilde q)\left[\frac{q^\beta  
 q^{\beta'}}{\tilde q^{2}}({\cal P}_S)_{\beta'\alpha}
-\frac{1}{3}\delta_{_{2S,3}}\, g^{\beta}_{\;\;\alpha}\right] \right.
 \nonumber\\
 & \left.  \quad\times
\sum_{i=1}^2{\cal O}_i^\mu
\left(\gamma^{\alpha} -\frac{P_-^{\alpha}[\not\!P_+-M]}{mM+b}\right) \, 
\right\}
\,\psi_\Delta^{D\,2S}\psi_N^S
. \label{eq:514}
\end{align}
This will now be evaluated for the two cases of interest.

\subsubsection{Nucleon(S) $\to$ $\Delta$(D3)}

The term in round brackets in Eq.~(\ref{eq:511}) commutes with ${\cal O}_2$ (an identity operator on the Dirac space), and hence, for the transition to the spin 3/2 core state (D3) with ${\cal P}_S={\cal P}_{3/2}$ gives zero  (this is the $\rho$ term discussed above).  Commuting the term in round brackets through ${\cal O}_1$, letting $\not\!P_+\to M$ when it operates to the left, gives
\bea
&&\gamma^\mu \left(\gamma^{\alpha} -\frac{P_-^{\alpha}[\not\!P_+-M]}{mM+b}\right)
\nonumber\\
&&\qquad= 2g^{\alpha\mu} -\gamma^\alpha\gamma^\mu-\left(\frac{P_-^{\alpha}[2P_+^\mu-2M\gamma^\mu]}{mM+b}\right).\qquad \label{eq515}
\eea
For the S=3/2 case under consideration, the $\gamma^\alpha\gamma^\mu$ terms vanishes, and combining this with the remaining terms gives
\begin{align}
\overline{\Gamma}_{D3}^{\beta\,\mu}(P,q)&=
-2\sqrt{3}\,f_v
\int_k
\left\{
\;b(\tilde k, \tilde q)\left[\frac{q^\beta  
 q^{\beta'}}{\tilde q^{2}} ({\cal P}_{3/2})_{\beta'\alpha}
-\frac{1}{3}  g^{\beta}_{\;\;\alpha}\right] \right.
 \nonumber\\
 &
\left.
\qquad\times \Bigg(g^{\alpha\mu} -\frac{P_-^{\alpha}[
P_+^\mu-M\gamma^\mu]}{mM+b} \Bigg)\, 
\right\} 
\,\psi_\Delta^{D3}\psi_N^S
. \label{eq516}
\end{align}
Now, we know that the terms proportional to $q^\mu$ can be ignored (they determine $G_3$ which we already know is just right to give a gauge invariant result, but otherwise play no role in the calculation).  Furthermore, we already know that $G_C^*=0$, and hence the value of $G_2$ must be fixed in terms of $G_1$ and $G_4$ through Eq.~(\ref{eqGEGM}), so we need not calculate it explicitly.  This leaves only $G_1$ and $G_4$, whose values can be extracted from (\ref{eq516}) by separating out the terms dependent on $g^{\beta\mu}$ and $q^\beta\gamma^\mu$.  This leads to 
\bea
G_1&=&0\nonumber\\
G_4&=&-\frac{2}{\sqrt{3}}\,f_v\, {\cal I}_{D3}
\eea
where the overlap integral ${\cal I}_{D3}$ is
\be
{\cal I}_{D3}=
\int_k b(\tilde k,\tilde q)\,\psi_\Delta^{D3}(P_+,k) \psi_N^S(P_-,k).
\ee
From Eqs.~(\ref{eqGM1}) and (\ref{eqGE1}) we obtain 
\ba
& &
G_M^{D3} (Q^2)=-\frac{4}{3 \sqrt{3}} 
\frac{m}{M+m} f_v \,{\cal I}_{D3} \\
& &
G_E^{D3} (Q^2)=-\frac{4}{3 \sqrt{3}} 
\frac{m}{M+m} f_v \,{\cal I}_{D3} \\
& &
G_C^{D3} (Q^2)=0.
\ea

Although formally different from the integral 
involved in the S-state transition, it can be shown 
that the integral ${\cal I}_{D3}$  goes with $1/Q^4$ for large $Q^2$. 
The proof follows the lines presented 
for case I of Appendix G from Ref.~\cite{NDelta}.
As consequence $G_M^{D3}=G_E^{D3} \sim 1/Q^4$.

\subsubsection{Nucleon(S) $\to$ $\Delta$(D1)}

For the D1 transition the $\rho$ term is no longer zero, 
and using  Eq.~(\ref{eq:514}), the property 
of the $S=1/2$ projection operator, 
and the definition of ${\cal O}_2$ gives
\begin{align}
\overline{\Gamma}_{D1}^{\beta\,\mu}(P,q)\Big|_{\rho}&
=\sqrt{3}\,f_{2-} \int_k 
\left\{
b(\tilde k, \tilde q)\, q^\beta  \frac{P^\mu}{m} \right. \nonumber\\
& \left.
\qquad\qquad\times\frac{1}{\tilde q^{2}}\;
 q^{\beta'} ({\cal P}_{1/2})_{\beta'\alpha}\,
\gamma^{\alpha} 
\right\}
\,\psi_\Delta^{D1}\psi_N^S
, \label{}
\end{align}
which contributes only to $G_2$
\bea
G_2\Big|_{\rho}=-\frac{2\sqrt{3}M}{m\,d_-}\,f_{2-}\; 
{\cal I}_{D1}\label{eqGrho}
\eea
 where the D1 overlap integral is 
\be
{\cal I}_{D1}=  
\int_k b(\tilde k,\tilde q)\, 
\psi_\Delta^{D1} (P_+,k) \psi_N^S(P_-,k).
\ee
Comparing this calculation with Eq.~(\ref{eqJgen}), 
and using the connection $j_2 \to C_0\,f_{2-}$, 
gives an explicit expression for $\rho_{\frac12\frac12}$
\bea
\rho_{_{\frac12\frac12}}(Q^2)&=&{\cal R}_{_{\frac12\frac12}}\frac{2M}{\sqrt{3}\,d_-}\; {\cal I}_{D1} 
\nonumber\\
&=& \frac1{3C_0}\sqrt{\frac{d_+}{2Mm}}\; {\cal I}_{D1}.\quad
\label{eq:rhod1}
\eea

Next, using Eq.~(\ref{eq515}) for $S=1/2$ case 
(where the $\gamma^\alpha\gamma^\mu$ term does not vanish), 
the ${\cal O}_1$ term for the D1 transition is  
\begin{align}
&\overline{\Gamma}_{D1}^{\beta\,\mu}(P,q)\Big|_{{\cal O}_1}=
-\sqrt{3}\,f_v \int_k 
\left\{
b(\tilde k, \tilde q)\left[\frac{q^\beta  
 q^{\beta'}}{\tilde q^{2}} ({\cal P}_{1/2})_{\beta'\alpha}\right]
\right.
 \nonumber\\
 &
\left.
\quad\times \Bigg(2g^{\alpha\mu}-\gamma^\alpha\gamma^\mu -\frac{P_-^{\alpha}[2P_+^\mu-2M\gamma^\mu]}{mM+b} \Bigg)\, 
\right\}
\,\psi_\Delta^{D1}\psi_N^S
. \label{}
\end{align}
From this we must extract the contributions to $G_1$, $G_2$, and $G_4$ [again ignoring $G_3$ which, using the modified current (\ref{eqJI2}), will be given by the gauge invariant condition].  It is easy to see that $G_4=0$, and
\bea
G_1&=&\frac{2M}{\sqrt{3}\,d_+}\,f_v\,{\cal I}_{D1} \nonumber\\
G_2\Big|_{{\cal O}_1}&=&\frac{8M (2m+M)}{\sqrt{3}\,d_+d_-}\,f_v {\cal I}_{D1}\, . \label{eq526}
\eea
These contributions combine with (\ref{eqGrho}) to give
\ba
& &G_M^{D1} (Q^2)=   
\frac{2}{3 \sqrt{3}}
   \frac{m}{M+m}\, f_v \, {\cal I}_{D1} \nonumber \\
& &G_E^{D1} (Q^2)=   -\frac{2}{3 \sqrt{3}} 
\frac{m}{M+m} \,f_v\, {\cal I}_{D1} \nonumber\\
%& &
%G_C^{D1} (Q^2)= \frac{4Mm}{ \sqrt{3}} \left[ f_{1-} 
%- \frac{Q^2}{2m(M+m)}f_{2-} 
%\right] {\cal I}_{D1}, \nonumber  \\
%& &\label{eqGC2}
& &
G_C^{D1} (Q^2)= \frac{4mM}{\sqrt{3}\;Q^2} \, f_{C} 
\,{\cal I}_{D1}   \label{eqGC2}
\ea
where $f_C$ is the analogue of (\ref{eq:jc})
\bea
f_C=f_{1-}-\frac{Q^2}{2m(M+m)}\,f_{2-}\, . \label{eq:fc}
\eea
Note that the expression (\ref{eqGC2}) for $G_C^{D1} (Q^2)$ is consistent with (\ref{eqJ02aa}) if we use the expression (\ref{eq:rhod1}) and the connection 
$j_C \to C_0\,f_C$.

Finally, as we have already discussed, the possible singularity in $G_C^*$ must be canceled by imposing the requirement 
\bea
\lim_{Q^2\to0} {\cal I}_{D1}\to A\; Q^2
\label{eq528}
\eea
where $A$ is a constant.  This constraint predicts that the D1 contributions to the magnetic and electric form factors will be zero at $Q^2=0$.
However, the constant $A$ in the limit (\ref{eq528}) will in general be nonzero, predicting that $G_C^\ast$ is finite as $Q^2\to0$.

For large $Q^2$, we can write ${\cal I}_{D1}$ 
as a difference of two integrals of the 
type ${\cal I}_{D3}$ with different coefficients.
Hence ${\cal I}_{D1}$ goes like $1/Q^4$, 
which gives a $1/Q^4$ behavior for $G_M^{D1}, G_E^{D1}$
and $G_C^{D1} \sim 1/Q^6$
(because $f_C \to$ constant as $Q^2\to\infty$).

In the overall, the asymptotic expression for the 
form factors   
are consistent with pQCD \cite{Carlson}. 

\subsection{Sum of all valence contributions}
%\subsection{Final result}
\label{secFFsum}

Considering the sum of all valence quark contributions,
we obtain the contribution of the quark core,
which we denominate by 'bare' (B) contribution: 
\ba
G_M^B(Q^2)&=&
N\left[G_M^S + a \,G_M^{D3} + b \,G_M^{D1}\right] 
\label{eqGMB}\\
G_E^B(Q^2)&=&
N\left[a\, G_M^{D3} - b\, G_M^{D1} \right] 
\label{eqGEB}\\
G_C^B(Q^2)&=&
N b\,  G_C^{D1}, 
\label{eqGCB}
\ea
where we used the relations between the 
electrical and magnetic components for each state.
Note that there are only two 
contributions for $G_E^B$, 
and one of them ($G_M^{D1}$) is zero for $Q^2=0$.
As for $G_C^B$ there is only the D1 state contribution.

For completeness we mention here that the nucleon could also have a D-state.  However, the nucleon (with total angular momentum $J=1/2$) 
can only have the D-state with core spin 3/2. 
This nucleon D-state can be build using  
the ideas presented in the previous sections
and leads to an additional contribution to  $G_C^\ast$.
We have not considered such a D-state admixture in this paper because the nucleon form factors can be well described at low $Q^2$ \cite{Nucleon} without including it.

\section{Pion Cloud contribution to the form factors}
\label{secPionCloud}

%\subsection{Pion Cloud parametrization}

The previous section  presented the 
contribution for 
the form factors from  the 
photon-quark interaction in relativistic impulse approximation, 
and within the spectator
theory.
%The 'bare' contribution from the valence 
%quark were presented in Sec.\ref{secFFsum}.
%
But the description of the 
electromagnetic $N \Delta$ transition 
requires also the presence of non-valence 
degrees of freedom, which may involve 
two-body currents and/or sea quark contributions
-- dominated by virtual pion states, the pion cloud effects.

In the language of the dynamical models, 
where the hadronic interactions are 
described in terms of a baryon core  
which  interacts with mesonic fields, 
a transition form factor can be separated 
into two terms \cite{Diaz06a,Pascalutsa06b,Burkert04}:
the contribution of the quark core,
or Bare contribution, and the 
contribution from the pion cloud:
\be
G_\alpha^\ast (Q^2)=
G_\alpha^B (Q^2) + G_\alpha^\pi (Q^2),
\label{eqGalpha}
\ee
where $\alpha$ holds for M,E,C 
and $G_\alpha^\pi$ denotes the corresponding 
mechanisms involving at least one intermediate pion state.
This contribution is  
related with the long range interaction, while
$G_\alpha^B$ contains the short range 
physics \cite{Diaz06a}
parametrized by the baryon wave functions.
The decomposition (\ref{eqGalpha}) was 
also considered in Ref.~\cite{Buchmann07a}. 
%Note that this decomposition scheme is model dependent, 
%since the separation bewteen the
%background and the resonant amplitudes 
%is not unique \cite{Bernstein03,Bernstein07,Diaz06a,Meissner07}.

Note that this scheme is model dependent, 
because the decomposition in 
background and resonances amplitudes 
is not unique \cite{Bernstein03,Bernstein07,Diaz06a,Meissner07}.
However, once established the pion production mechanism ($\pi$NN  amplitude), 
we can split $G_\alpha^\ast$ in two contributions 
in a given formalism.

Although our main goal here  
is to study the D-state effects 
in  the core valence quark wave function, even a qualitative estimate of the D-states 
effects requires a simulation of the 
pion cloud effects.
An effective parametrization 
of the pion cloud in $G_M^\ast$ was 
already introduced in a previous work \cite{NDelta}.
For $G_E^\ast$ and $G_C^\ast$ we consider in the present work  
the parametrization introduced in 
references  \cite{Pascalutsa07a,BuchmannEtAl,Buchmann01,Buchmann04}, 
which we will
sketch now.

\subsection{Pion cloud parametrization of $G_C^\ast$}

In a pure SU(6) model the neutron electric
form factor $G_{En}$  would be identically zero 
and the multipoles E2 and C2 in the 
$\gamma N \to \Delta$ transition negligible.
In the real world $G_{En}$ is small but non-zero.

Considering a constituent quark model 
with a confining harmonic oscillator 
potential with also pion- and gluon- exchange 
between quarks, Buchmann \cite{BuchmannEtAl,Buchmann07a} concluded that
the $G_{En}$ data can be explained considering  a two-quark current, with a  
quark-antiquark pair interacting with the external photon.
In this description the neutron spatial extension, 
expressed in term of its radius, can be written as  
\be
r_n^2= - \frac{M^2-m^2}{m} b_q^2,
\ee
where $b_q$ is the quark core radius 
(oscillator parameter).
For 
the experimental result: $r_n^2 \simeq -0.113 \;\mbox{fm}^2$, 
we can estimate $b_q \approx 0.6$ fm. 
Within the same formalism, one concludes \cite{BuchmannEtAl} that
\be
G_C^\ast(0)= -\sqrt{\frac{2m}{M}} M m \frac{r_n^2}{6}.
\ee  
As, for low $Q^2$, we can write for $G_{En}$ 
\be
G_{En}(Q^2) \simeq - Q^2  \frac{r_n^2}{6},
\label{eqGEnR}
\ee
and we obtain, for small $Q^2$:
\be
G_C^\ast (Q^2)=
\sqrt{\frac{2m}{M}} M m \frac{G_{En}(Q^2)}{Q^2}.
\label{eqGCpion}
\ee
The relation (\ref{eqGCpion}) can alternatively also 
be constructed from relations 
between the nucleon and nucleon to $\Delta$ transition
magnetic moment, in the 
large $N_c$ limit \cite{Pascalutsa07a}, 
for low $Q^2$ ($Q^2 << 1$ GeV$^2$).  

Following Buchmann again,  
from a different perspective \cite{Buchmann07a}, the 
nucleon form factors can be described by a symmetric quark core distribution 
plus an asymmetric pion cloud around the inner core.
Considering the proton electrical form factor 
in particular, we can write
\be
G_{Ep}(q^2) = G_{Ep^\prime}(Q^2) + G_{Ep}^{\pi} (Q^2),
\label{eqGEp}
\ee
where $ G_{Ep^\prime}(Q^2)$ is the 'bare' proton 
charge form factor, and 
$G_{Ep}^{\pi} (Q^2)$ 
is the contribution 
due to the pion cloud.
In the same picture the 
neutron electric form factor is however just given 
by the pion cloud and we may write
\be
G_{En}(Q^2) = -G_{Ep}^{\pi} (Q^2),
\label{eqGEn}
\ee
since the charge distribution in the neutron 'bare' core is zero.
In the $Q^2=0$ limit  Eqs.~(\ref{eqGEp}) and (\ref{eqGEn})
are directly related with the nucleon radii.
From Eq.~(\ref{eqGEnR}) we obtain $G_{En}(0) \sim r_n^2 \approx -0.113$ fm$^2$.
As for $G_{Ep}$ we may write $G_{Ep}(Q^2) \simeq 1- r_p^2 \frac{Q^2}{6}$, 
where $r_p^2$ is the proton electrical squared radius.
Now, $r_p^2$ can be decomposed as 
$r_{p^\prime}^2-r_n^2 \approx 0.78$ fm$^2$, 
where $r_{p^\prime}^2 \approx 0.67$ fm$^2$ 
represents the radius of the bare proton,
the size of the proton being increased by the pion cloud.

\begin{figure}[t]
%\vspace{1.0cm}
\centerline{
\mbox{
\includegraphics[width=7.0cm]{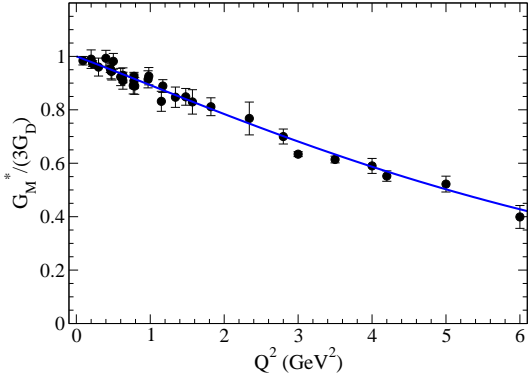} }}
%\vspace{-2cm}
\caption{Comparing the $G_M^\ast$ data with 
the parametrization of Eq.\ (\ref{eqGMest}).}
\label{GMest}
\end{figure} 

As $G_{En}$ is determined by 
pure pion cloud effects,
we conclude that $G_C^\ast$ 
(\ref{eqGCpion}) is the result of pion cloud effects
(or equivalently, the Coulomb quadrupole form factor 
in the $\gamma N \to \Delta$ transition 
would be zero, for the case of no pion cloud effects). The  previous derivation assumes 
no contribution from the inner core (symmetric distribution 
in the core).
This assumption is not valid in general 
but can be a good approximation for a small  D-state admixture.
We will therefore use
\be
G_C^\pi (Q^2)=
\sqrt{\frac{2m}{M}} M m \frac{G_{En}(Q^2)}{Q^2},
\label{eqGCpion2}
\ee
to represent the contribution of the pion cloud 
for $G_C^\ast$.

\begin{figure}
%\vspace{1.0cm}
\centerline{
\mbox{
\includegraphics[width=6.9cm]{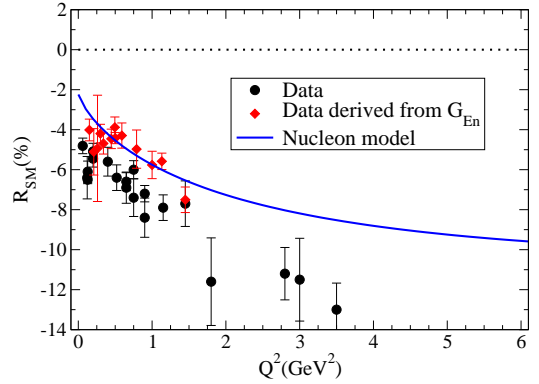} }}
%\vspace{-2cm}
\caption{
Comparing the $R_{SM}$ data with the prediction 
of Eq.~(\ref{eqGCpion}) using  the neutron electrical 
form factor data of Ref.~\cite{Nucleon}.
The nucleon model line corresponds to the
model of Ref.~\cite{Nucleon}.
The nucleon wave function 
parameters are presented in table \ref{Nucleon_table}.
}
\label{fig4}
\end{figure}

To check the consistency of this assumption, and before using it together
with the bare model built here,
we compare the $R_{SM}$ data with the 
results extracted from the electrical 
form factor data using the parametrization (\ref{eqGCpion}).
To estimate $G_M^\ast(Q^2)$ at the respective momentum 
$Q^2$, we consider the simple 
phenomenological parametrization of  Ref.~\cite{Gail06}:
\be
G_M^\ast(Q^2) =
3 G_D \exp(-0.21 Q^2) \sqrt{1+ \frac{Q^2}{(M+m)^2}}.
\label{eqGMest}
\ee
The quality of this parametrization for $G_{M}^\ast$ is 
presented in Fig.~\ref{GMest}.
The results are presented in Fig.~\ref{fig4}, where we calculated $G_{En}$
from our spectator constituent quark model.
Although according to 
Eq.~(\ref{eqGCpion}) the pion cloud contribution to $G_C^\ast$
decreases with $Q^2$, its effect is not 
observed in  the figure, due to 
the kinematic factor 
$\frac{|{\bf q}|}{2M}$ present in $R_{SM}$.

%From the figure we can conclude 
%that the assumption is consistent for $Q^2 < 0.5$ GeV$^2$ 
%but the pion cloud parametrization ($G_{En}$) 
%underpredict $R_{SM}$ for  $Q^2 > 0.5$ GeV$^2$.

Due to the nature of the derivation of Eq. (\ref{eqGCpion2}) 
(large $N_c$ limit and 
$Q^2 \sim 0$) we cannot say for sure 
whether or not the  discrepancies in Fig.\ \ref{fig4}  
are the result of the crude estimation 
(${\cal O}(1/N_c^2)$ correction to the large $N_c$ limit)  
or the result of neglecting the bare quark contribution.
Reference \cite{Buchmann01} estimates 
the D-state effects from this one-body current 
to be  20\% to the final result.

\subsection{Pion cloud parametrization of $G_E^\ast$}

\begin{figure}
%\vspace{1.0cm}
\centerline{
\mbox{
\includegraphics[width=6.9cm]{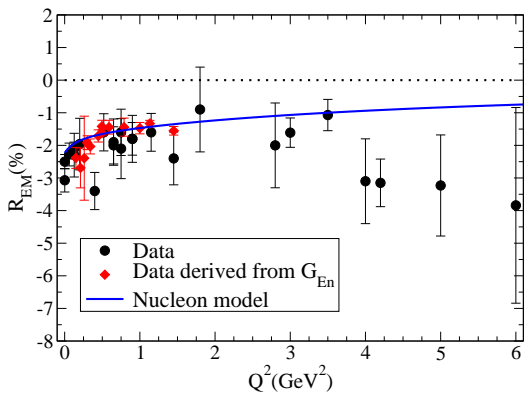} }}
%\vspace{-2cm}
\caption{ 
Comparing the $R_{EM}$ data with the prediction 
of Eq.~(\ref{eqGEpion}) using  the neutron electrical 
form factor data of Ref.~\cite{Nucleon}.
The nucleon model line corresponds to the model of Ref.~\cite{Nucleon}.
The nucleon wave function 
parameters are presented in table \ref{Nucleon_table}. }
\label{fig5x}
\end{figure}

Considering the large $N_c$ limit,  
Pascalutsa and Vanderhaghen \cite{Pascalutsa07a}
related $G_C^\ast$ and $G_E^\ast$ 
at the photon point ($Q^2=0$)
\be
G_C^\ast(0)= \frac{4M^2}{M^2-m^2} G_E^\ast (0).
\label{eqLongW}
\ee
Using the relation (\ref{eqGCpion}) between 
$G_C^\ast$ and $G_{En}$, and extending the 
results for finite $Q^2$, one has
\cite{Pascalutsa07a}:
\be
G_E^\ast (Q^2)= 
\left(\frac{m}{M}\right)^{3/2} \frac{M^2-m^2}{2\sqrt{2}} 
\frac{G_{En} (Q^2)}{Q^2}.
\label{eqGEpion}
\ee
This result was derived in  Ref.~\cite{Pascalutsa07a}, 
in the $Q^2=0$ limit, and 
must be restricted to low $Q^2$ ($Q^2 <<  $ 1 GeV$^2$).
The comparison between the $R_{EM}$ data and the 
predictions from Eq.~(\ref{eqGEpion}) using the 
$G_{En}$ data is presented in Fig.~\ref{fig5x}. 

In the $G_C^\ast$ case, the exact SU(6) 
symmetry would imply $G_{En}\equiv 0$ 
and there is no contribution for the 
electric quadrupole. 
In contrast,
one cannot conclude that  Eq.~(\ref{eqGEpion}) 
results uniquely from pure pion cloud effects.
In the large $N_c$ analysis, both 
$G_E^\ast $ and $G_C^\ast$ are 
%of the order 
${\cal O}(1/N_c^2)$, to be compared with 
$G_M^\ast={\cal O}(N_c^0)$,
which is estimated in terms of the 
magnetic form factor of the neutron \cite{Pascalutsa06b,Buchmann07a}.
In that limit the  valence quark core is 
dominant, but the next order 
correction can be originated by pion cloud effects 
or by
angular momentum excitation of a quark.
But, because $G_E^\ast$ can be written 
in terms of $r_n^2$ (or $G_{En}(Q^2)$ for $Q^2 \sim 0$), 
%we  will use the same argument already used for 
%$G_C^\ast$ 
we will take (\ref{eqGEpion}) 
as the pion cloud contribution for $G_E^\ast$
for low $Q^2$, neglecting next order corrections ${\cal O}(1/N_c^3)$)
in the large $N_c$ limit,
\be
G_E^\pi (Q^2)= 
\left(\frac{m}{M}\right)^{3/2} \frac{M^2-m^2}{2\sqrt{2}} 
\frac{G_{En} (Q^2)}{Q^2}.
\label{eqGEpion2}
\ee
Reference \cite{BuchmannEtAl} estimates 
the contributions due to the quark-antiquark states $G_E^\pi$
to be 88\% of $G_E^\ast$ for $Q^2=0$.

The relation between $G_E^\ast$ and $G_C^\ast$ represented 
in Eq.~(\ref{eqLongW})
is also known as the long wavelength limit for the ratio 
$G_C^\ast(0)/G_E^\ast(0))$.
It is the result of the conditions: $Q^2 << M^2-m^2 << M^2,m^2$, 
like in large $N_c$ limit, where $M-m={\cal O}(1/N_c)$.
Equation (\ref{eqLongW}) is also used to relate 
the electrical 
and Coulomb bare quadrupoles in the SL \cite{Diaz06a}
and DMT \cite{Kamalov99} models.

A direct consequence of  (\ref{eqLongW}), if $G_C^\pi$ and $G_E^\pi$ 
are the only contribution for the respective form factors, is that
\cite{Pascalutsa07a}  
\be
R_{EM} (0) =R_{SM}(0).
\ee

Note that the pion cloud contributions 
(\ref{eqGCpion2}) and  (\ref{eqGEpion2})
for $G_C^\ast$ and $G_E^\ast$ respectively, goes with $1/Q^6$ 
for large $Q^2$, competing with the bare 
contributions ($1/Q^6$ and $1/Q^4$ respectively).
[Assuming as in Ref.~\cite{Nucleon} 
that $G_{En} \sim 1/Q^4$].
As a consequence, the pion cloud 
contribution does not change 
the asymptotic behavior 
derived for $G_E^B$ and $G_B^C$.
We need to have in mind, however, 
that the results 
for $G_C^\pi$ and $G_E^\pi$ 
are derived under the assumption 
that $Q^2$ is small.
Buchmann \cite{Buchmann04,Buchmann07a} 
argues that nevertheless, the pion cloud 
description for $G_C^\ast$ can be extended 
also to the  intermediate $Q^2$ region ($Q^2 \sim 4$ GeV$^2$).

With the parametrization of the pion cloud 
mechanisms using the Equations 
(\ref{eqGCpion2}) and (\ref{eqGEpion2})
we preserve the covariance of our calculation
because $G_{En}$ is evaluated 
using a spectator model \cite{Nucleon}.

\section{Results}
\label{secResults}

In this section we present the numerical results of 
our model to the $\gamma N \to \Delta$ transition.
For the quark current 
we adopted the quark form factors 
from Ref.~\cite{Nucleon} 
based on a
vector dominance model (VDM) parametrization:
\ba
f_{1\pm} (Q^2) &= &
\lambda + \frac{(1-\lambda)}{1+Q^2/m_v^2} +
\frac{c_{\pm} Q^2/M_h^2}{\left(1 + Q^2/M_h^2 \right)^2} \qquad
\nonumber\\ 
f_{2\pm} (Q^2) &= &
\kappa_{\pm} 
\left\{ 
\frac{d_\pm}{1+Q^2/m_v^2} 
+
\frac{(1-d_{\pm})}{1+Q^2/M_h^2} 
\right\}\, . \label{eqf1m}
\ea
In these expressions $m_v$ and $M_h$ are  
the masses of the vectorial mesons.
The lower mass, $m_v= m_\rho$ (or $m_\omega$), 
describes the two pion resonance 
(three pion resonance) effect 
and $M_h =2m$, 
takes account of all the larger mass resonances.
The parameter $\lambda$ was  
adjusted to give  the correct quark density number 
in deep inelastic scattering  \cite{Nucleon,NDelta}. 
All the other parameters are presented in Table
\ref{Nucleon_table}.

We will divide this section into two subsections.
First we consider the effects of the valence 
quarks. 
In particular we test whether the bare contributions
alone calculated as explained in Sec.~\ref{secFFval} 
can describe the experimental data.
In the second subsection we add the effects of the 
sea quarks (pion cloud effects), 
with the phenomenological, parameter free, 
description of the pion cloud presented in the Sec.~\ref{secPionCloud}.

\begin{table}[t]
\begin{center}
\begin{tabular}{c c c c  c}
$\beta_1$, $\beta_2$ & $c_+, c_-$ & 
$d_+,d_-$ & $\lambda, m_s/m$ \\
\hline
0.049      &  4.16  & -0.686 &  1.21   \\   
0.717      &  1.16  & -0.686 &  0.87 \\ 
\hline
\end{tabular}
\end{center}
\caption{Parameters of the nucleon wave function 
($\beta_1,\beta_2$) and quark form factors 
corresponding to the model II of Ref.\  \cite{Nucleon}.
In each case we kept $\kappa_+= 1.639$ and 
$\kappa_-=1.823$ in order to reproduce  
the nucleon magnetic moments exactly.} 
\label{Nucleon_table}
\end{table}

\subsection{Valence quark contributions only: Models 1  - 3}
\label{secResA}

\begin{figure}[t]
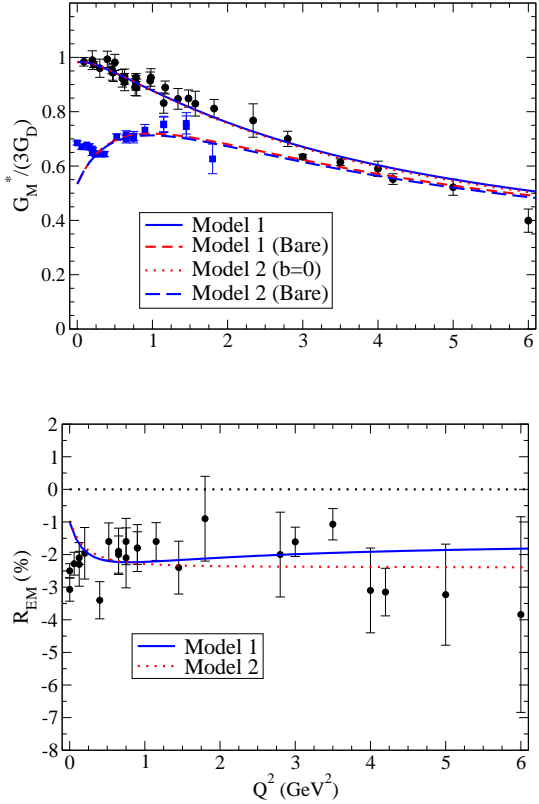

%\vspace{1.0cm}
\centerline{
\mbox{
\includegraphics[width=7.1cm]{GMmI1} }}
\vspace{.85cm}
\centerline{
\mbox{
\includegraphics[width=6.9cm]{REMmI1} }}
%\vspace{.5cm}
%\centerline{
%\mbox{
%\includegraphics[width=7.0cm]{RSMmI1.eps} }}
\caption{Models 1 and 2.
$G_M^\ast$ data from from CLAS/Jlab \cite{CLAS02,CLAS06}, 
DESY \cite{Bartel68} and SLAC \cite{Stein75}. 
$R_{EM}$ data from MAMI \cite{MAMI,Stave06a},
LEGS \cite{LEGS}, MIT-Bates \cite{Bates}
and Jlab \cite{CLAS02,CLAS06}.
The ``bare'' data for $G_M^B$, shown in the top panel,
were extracted from the SL analysis, Ref.~\cite{Diaz06a}.  
For the fit we doubled the bare data error bars shown
in the figure to constrain $G_M^B$,
but the extra $\chi^2$ that results from the fit of 
$G_M^B$ to bare data is not included in any of the $\chi^2$  
reported in this paper.  }
\label{fig1}
\end{figure} 

With only valence quark
degrees of freedom the $N\Delta$ electromagnetic transition 
form factors are described by Eqs.~(\ref{eqGMB})-(\ref{eqGCB}).
The free parameters of our model are
the admixture coefficients $a$, $b$  and 
the momentum range parameters of 
the scalar wave functions (\ref{eqP1})-(\ref{eqP3}).
In a previous work we adjusted the S-state $\Delta$ 
wave function to the $G_M^\ast$ data 
considering also an effective pion cloud 
contribution \cite{NDelta} as
\be
G_M^\ast(Q^2) = 
G_M^B(Q^2) + G_M^\pi(Q^2),
\label{eqGMs}
\ee
where $G_M^B$ is the contribution of 
the quark core and $G_M^\pi$ the pion cloud effects, parametrized by
\be
G_M^\pi (Q^2) = \lambda_\pi 
\left(\frac{\Lambda_\pi^2}{\Lambda_\pi^2 +Q^2} \right)^2 (3G_D),
\label{eqGMpi}
\ee
where $G_D=\left(1+Q^2/0.71\right)^{-2}$ 
is the nucleon dipole form factor, $\Lambda_\pi$ 
a cut-off and $\lambda_\pi$ a coefficient the 
defines the intensity of the pion cloud effect.
The factor $3$ was included for convenience:
when $Q^2=0$, $G_M^\pi(0)/G_M^\ast(0)=\lambda_\pi$, then 
$\lambda_\pi$ measures the fraction of pion cloud 
($G_M^\ast(0) \approx 3$). 
The parametrization (\ref{eqGMpi}) simulates the main 
features of the pion cloud mechanism:
significant contribution for $Q^2=0$; 
falloff with increasing $Q^2$.
For more details see Ref.~\cite{NDelta}.

Here we extend the predictions to 
the subleading quadrupole form factors 
$G_E^\ast$ and $G_C^\ast$ 
expressed in the ratios $R_{EM}$ and $R_{SM}$ 
defined respectively by Eqs.~(\ref{eqREM})-(\ref{eqRSM}).
We kept the parametrization (\ref{eqGMs})-(\ref{eqGMpi})
for $G_M^\ast$,
however $G_M^B$ is no longer 
determined only by the $\Delta$ S-state, 
but now also includes contributions of both of the D-states.
For this reason the parameters originally fixed in 
the S-state fit are now readjusted.

We considered the $G_M^\ast$ data from CLAS/Jlab \cite{CLAS02,CLAS06}, 
DESY \cite{Bartel68} and SLAC \cite{Stein75}.
For the electromagnetic 
ratios  $R_{EM}$ and $R_{SM}$ we use the 
data from MAMI \cite{MAMI,Stave06a},
LEGS \cite{LEGS}, MIT-Bates \cite{Bates}
and Jlab \cite{CLAS02,CLAS06}.
Although there is no inconsistency in  the $G_M^\ast$ data,
there is some ambiguity in 
the $R_{EM}$ and $R_{SM}$ data, dependent on the analysis.
For the form factor information 
to be  extracted one uses data for the pion 
photoproduction reaction cross sections. 
Those cross sections are interpreted in terms 
of an amplitude that includes a background and a 
resonant contribution.
In the process, the extraction of the multipoles depends  
on assumptions for the background 
and resonance parametrization.
The multipole resonant amplitudes 
are then varied to fit the cross section data
\cite{Bernstein03}.
Kamalov {\it et al.}~\cite{Kamalov01} presented a re-analysis 
of the CLAS-2002 data \cite{CLAS02} with significant differences 
from the original data. 
Similarly the CLAS-2006 $R_{SM}$ data \cite{CLAS06}  
for $Q^2 \ge 3$ GeV$^2$
shows a dependence on $Q^2$ different
from the recent MAID analysis \cite{Drechsel07} of the same data. 
The $R_{EM}$ analysis from 
Arndt {\it et al.}~\cite{Arndt07} is in contradiction 
with all the published results.
More recently Stave \cite{Stave08} showed 
that there is a significant discrepancies
in the extraction of E2 and C2 from the 
data using different reaction models like SL and DMT
in the region $Q^2 < 1$ GeV$^2$.
[This discrepancy can be reduced 
by refitting the models 
within  the range $Q^2 < 1$ GeV$^2$ only, which however prevents 
the range of the application of the models for higher $Q^2$ regions.]

\begin{figure}[b,t]
%\vspace{1.0cm}
\centerline{
\mbox{
\includegraphics[width=7.0cm]{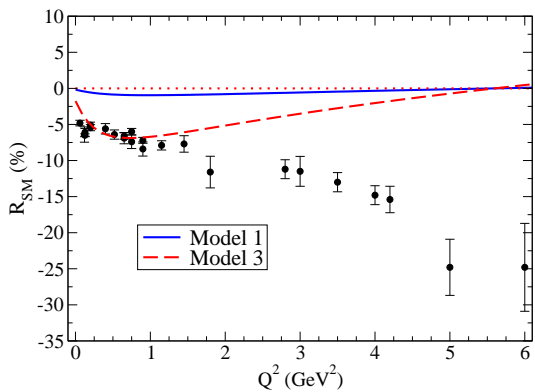} }}
%\vspace{-2cm}
\caption{$R_{SM}$ from models 1 and 3.
Data from MAMI \cite{MAMI,Stave06a},
LEGS \cite{LEGS}, MIT-Bates \cite{Bates}
and Jlab \cite{CLAS02,CLAS06}.}
\label{fig1b}
\end{figure}

As a first step, model 1 fits only the 
$G_M^\ast$ and  $R_{EM}$ data (using, as in Ref.\ \cite{NDelta}, 
the ``bare'' data extracted by the SL model \cite{Diaz06a}  
to constrain the bare form factor $G_M^B$).  
This fit (together with the  fit from model 2 
described below) is shown in Fig.~\ref{fig1}.  
The Coulomb form factor predicted by model 1 
(and not used in the fit) is shown in Fig.\  \ref{fig1b}. 

%%Table II
%\begin{widetext}
\begin{center}
\begin{table*}[t]
\begin{tabular}{c c c c c c c c c  c c}
Model   & $\lambda_\pi, \Lambda_\pi^2$ &
$\alpha_1, \alpha_2$ & & $\alpha_3, \alpha_4$ & &
$\lambda_{D1}, \alpha_5$ & & D3,D1 &  $\chi^2_{GM},\chi^2_{REM}$
& $\chi^2_{RSM}, \chi^2$ \\
\hline
 1    & 0.450 &  0.344  &  & 0.1956 & & 1.025  & & 8.15\% & 1.41 &  - \\
        & 1.46  &  0.344  &  & 0.1978 & & 0.1165 & & 0.17\%    & 4.39 &
{\bf 2.72} \\
\hline
 2    & 0.448 &  0.350  &  &    -   & &   -    & & 8.16\% & 1.21 &  - \\
        & 1.53  &  0.343  &  &    -   & & 0.0991 & &  -     & 4.90 & {\bf
2.83} \\
\hline
 3    & 0.479 &  0.343  &  & 0.1567 & & 1.0087 & & 8.50\%  & 3.33 &
11.84 \\
        & 1.30  &  0.350  &  & 0.1574 & & 0.2218 & & 15.2\%    & 3.80 &
{\bf 5.45} \\
\hline
 4    & 0.441 &  0.336  &  & 0.1089 & & 1.0094 & & 0.88\%  & 1.41 &
5.68 \\
        & 1.53  &  0.337  &  & 0.1094 & & 0.1880 & & 4.36\%  & 0.99 & {\bf
2.51} \\
\hline
\end{tabular}
\caption{
Model 1 fits $G_M^\ast$ and $R_{EM}$.
Model 2 fits the same quantities with $b=0$ (no D1 mixture).
Model 3 fits all variables but restricts $Q^2< 1.5$ GeV$^2$ for
$R_{SM}$.
Model 4 includes an effective pion cloud in both
$R_{EM}$ and $R_{SM}$ (for $Q^2 < 4.3$ GeV$^2$).
All models also fit $G_M^B$ to the bare data (as shown in the figures) but the extra $\chi^2$ that results from the fit to bare data is not included in the   $\chi^2$ reported in the last two columns. }
\label{tableMod1}
\end{table*}
\end{center}
%\end{widetext}

The parameters that were adjusted during the fits 
are shown in Table \ref{tableMod1}.
Although we did not fit the $R_{SM}$ data, 
the coefficient $b$ which determines the strength 
of the D1 state was adjusted during the fit.
[As emphasized in the previous sections, only the D1 state can generate
a non-vanishing $R_{SM}$, or  $G_C^\ast \ne 0$.]
As we see in Table  \ref{tableMod1} 
the best description of the $G_M^\ast$ and $R_{EM}$ data
requires a small admixture of the D1 state (0.2\%).
To check the sensitivity of the fit to the inclusion of the D1 state,
we considered also another fit forcing $b=0$. This defines model 2.
As Table \ref{tableMod1} 
shows, the admixture with $b\ne 0$ improves the description of the data only slightly 
($\chi^2$ of 2.72 versus 2.83), meaning that the role of the D1 state is not
decisive for $G_M^\ast$ and $R_{EM}$.

Figure~\ref{fig1b} shows the prediction of model 1  for 
$R_{SM}$ [the result for model 2  is zero].
We conclude 
that the $R_{SM}$ prediction from  model 1 is an 
order of magnitude smaller than the data.

The next step is to try to fit the $R_{SM}$ data as well, still using 
only the  valence quark degrees of freedom.  
However, because of the  zero in $f_C$, Eq.\ (\ref{eq:fc}), 
we also predict a zero in $G_C^\ast$, Eq.\ (\ref{eqGC2}). 
Using the parameters of Ref.~\cite{Nucleon}, 
$f_C$ passes through zero around $Q^2 \simeq 5.6$ GeV$^2$. This zero
is completely at odds with the data.
It is therefore impossible to fit $G_C^\ast$ over the entire $Q^2$ range, 
and 
at this stage we restrict  the fit to  
$R_{SM}$ to the  low momentum region $Q^2 < 1.5$ GeV$^2$.  
This fit defines model 3.

\begin{figure}
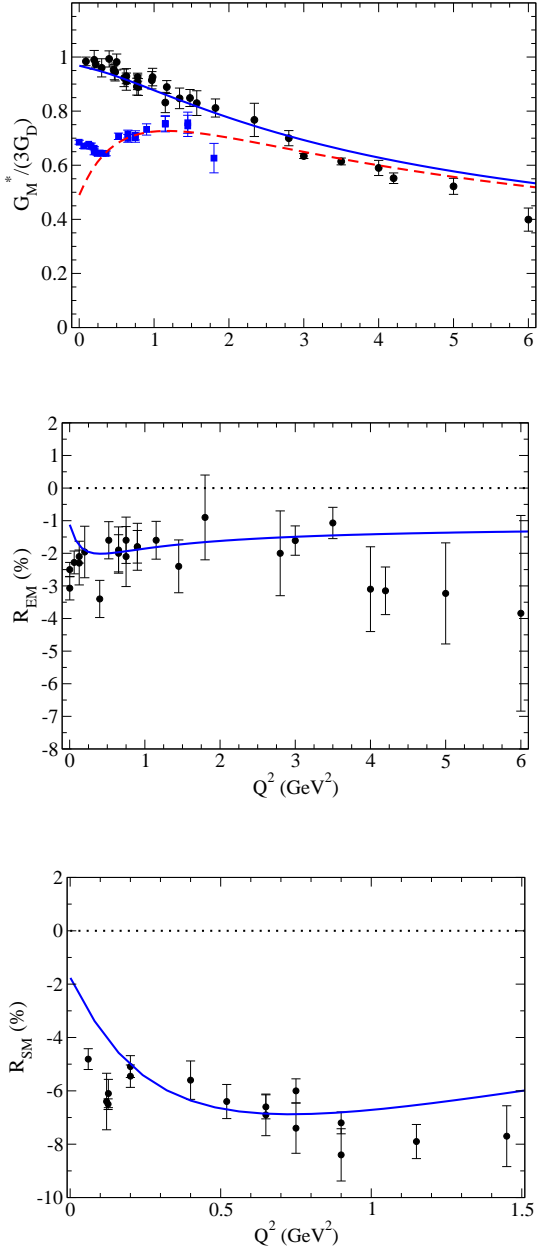

%\vspace{1.0cm}
\centerline{
\mbox{
\includegraphics[width=7.1cm]{GMmI3} }}
\vspace{.85cm}
\centerline{
\mbox{
\includegraphics[width=6.9cm]{REMmI3} }}
\vspace{.9cm}
\centerline{
\mbox{
\includegraphics[width=7.0cm]{RSMmI1d} }}
%\vspace{-2cm}
\caption{Model 3. 
Data from Figs.~\ref{fig1} and Fig.~\ref{fig1b}.}
\label{fig2}
\end{figure}

The results from model 3 are shown in Fig.~\ref{fig2}. 
In the last panel of the figure 
we show $R_{SM}$  for $Q^2 < 1.5$ GeV$^2$ only, the range  used in the fit.
Figure~\ref{fig1b} compares the results for $R_{SM}$ obtained from models 
1 and 3 over the entire $Q^2$ range. 
Note  the unavoidable zero for model 3 at $Q^2 \sim 5.6$ GeV$^2$.
The first conclusion from model 3 is 
that the fit  gives  $R_{SM}$ only within the region
$Q^2 < 1.5$ GeV$^2$, and that the fit is a poor one (high $\chi^2_{RSM}$).
Also, the quality of the description 
of the $G_M^\ast$ data
is affected, 
as we can conclude from Table  \ref{tableMod1},
by comparing   $\chi_{GM}^2$ obtained
in  model 3 with the corresponding values  obtained in model 1 and 2.
Note also that even the qualitative
description of $R_{SM}$ provided by model 3
requires a abnormally large admixture of the D1 states (15.2\%).
All these observations show the intrinsic limitations of 
a pure constituent quark model. They can be overcome 
by adding pion cloud effects, 
as motivated in the discussion in the previous section.

At low $Q^2$  the data tell us that 
$R_{SM}(0) \approx - 4\%$,
which is equivalent to 
$G_C^\ast (0) \approx 1.1$ 
[this follows from (\ref{eqRSM}) with $G_M^\ast (0) \simeq 3$].  
Without a pion cloud such a result can be obtained  
in this formalism only by requiring a large D1 admixture.
But even with a large D1 admixture, the valence quark contribution for $R_{SM}$ 
changes sign at about $Q^2 \sim 5.6$ GeV$^2$, and  we are led to conclude 
that the valence quark degrees of freedom 
are insufficient to explain the $G_C^\ast$ data for large $Q^2$.
This conclusion is consistent with both 
constituent quark models and the 
results from dynamical models.

Let us discuss now the numerical values of  the range parameters $\alpha_i$.
Remember that $\alpha_i$ can be interpreted 
as a Yukawa range parameter \cite{Nucleon,Gross04}, 
with a smaller $\alpha$ parameter representing a larger spatial range.
It is interesting  to see that in all the models with D-state components, $\alpha_1 \simeq
 \alpha_2$.
This finding differs from the results obtained previously for a $\Delta$ wave function with  only an S-wave component  \cite{NDelta}. 
For the pure S-state, $\alpha_1 \approx 0.3$ and $\alpha_2 \approx 0.4$.
Apparently,  the  introduction of the D-states sets a new long range scale,
 with $\alpha_i\sim 0.1$ to 0.2, for $i=3,4,5$.  These longer range scales
are also nicely consistent with the notion that the D-waves are peripheral.
Finally, it is worth mentioning 
that the similarity in the values for $\alpha_1, \alpha_2 \simeq 0.35$, 
suggests that the S-state effects are somehow 
model independent.
This feature suggests that, in the future,  
it might suffice to chose only 1 parameter to 
describe the S-state of the $\Delta$, showing that the improvement 
that accompanies the inclusion of the D-states is robust.

Summarizing:  a qualitative description 
of the $G_M^\ast$ and $G_E^\ast$ data can 
be obtained using  a $\Delta$ wave function composed predominantly of an S-state with admixtures of  D3 and D1 states. 
A fit based on a quark core requires an 8.2\% admixture of the  D3 state.
The inclusion of the D1 state is not at all necessary 
to explain the $G_M^\ast$ and $G_E^\ast$ data.
To explain the
$G_C^\ast$ data using only a quark core requires an unusual large 
admixture of the D1 state, and  
is only reasonable for low $Q^2$,  failing totally 
for $Q^2 > 2$ GeV$^2$ (see Fig.\ \ref{fig1b}).
The conclusion from models 1 - 3 is that 
a quark core only is not sufficient to explain the 
$\gamma N \to \Delta$ transition data, 
even when the $\Delta$ wave function  includes 
admixtures of D1 and D3 states.

\subsection{A mixed description: valence quarks and a pion cloud: Model 4   }
\label{secResD}

We now add a pion cloud contribution to the $G_E^*$ and $G_C^*$ form factors
\ba
& &G_E^\ast(Q^2) = 
G_E^B(Q^2) + G_E^\pi(Q^2)\nonumber\\
& &
G_C^\ast(Q^2) = 
G_C^B(Q^2) + G_C^\pi(Q^2),
\label{eqGCtotal} 
\ea
where the pion cloud contributions 
are taken from Eqs.~(\ref{eqGEpion2}) and  (\ref{eqGCpion2}), respectively. 
There are no adjustable parameters
in the pion cloud components.
For the neutron electric form factor $G_{En}(Q^2)$ we use 
model II of the Ref.~\cite{Nucleon} (see table \ref{Nucleon_table}).
The limit of validity of the pion cloud formulas %in the large $N_c$ limit
is restricted to low $Q^2$,
which led us to restrict our fit to $R_{SM}$ to 
$Q^2 < 4.3 $ GeV$^2$ region.
The bare contributions from valence quarks
come from (\ref{eqGEB})-(\ref{eqGCB}).
Then, $G_E^B$ is the result 
of the valence quark contribution 
involving the D3 and D1 states, 
and $G_C^B$ comes only from the D1 state.

\begin{figure}
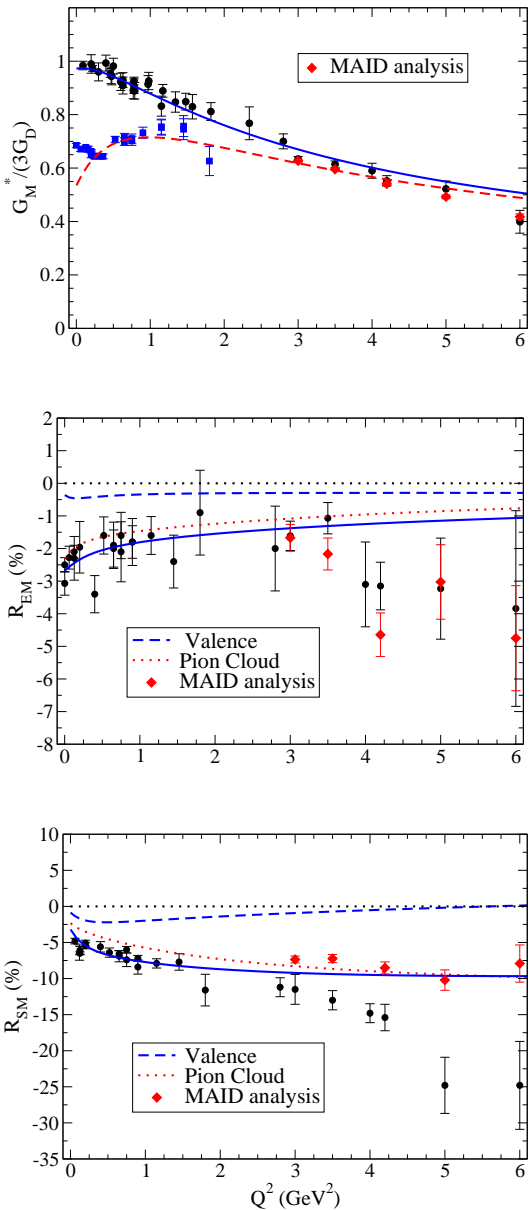

\centerline{
\mbox{
\includegraphics[width=7.0cm]{GMmI6} }}
\vspace{.78cm}
\centerline{
\mbox{
\includegraphics[width=6.9cm]{REMmI6} }}
\vspace{.78cm}
\centerline{
\mbox{
\includegraphics[width=7.0cm]{RSMmI6} }}
%\vspace{-2cm}
\caption{Model 4.
Data from Figs.~\ref{fig1}, Fig.~\ref{fig1b} and the recent 
MAID analysis 
\cite{Drechsel07}.}
\label{fig6}
\end{figure}

The results for model 4, the fit obtained 
using the pion cloud, are shown in Fig.~\ref{fig6}, 
with the parameters given in table \ref{tableMod1} 
(along with the results for all the other models).
The description of $G_C^\ast$ (and the  
corresponding ratio  $R_{SM}$) by model 4 for $Q^2>2$ GeV$^2$ 
favors the recent MAID analysis \cite{Drechsel07}
over the original JLab analysis \cite{CLAS06},
and its success or failure will ultimately 
depend on which of these analysis survives further study. 

By comparing the $\chi^2 $s in table \ref{tableMod1}  
for models 3 and 4, one concludes that 
the inclusion of the pion cloud for $G_E^\ast$
and $G_C^\ast$ does indeed improve significantly the 
simultaneous description of the data of these two  
more problematic observables.
In model  4 the contribution of the pion cloud at $Q^2=0$  
is 86.9\% for $G_E^\ast$, and  72.5\% for $G_C^\ast$.
The most important D state is the D1, with a small 
admixture of 4.4\% compared to a tiny admixture of D3 state of only 0.9\%.
As for the dominant $G_M^\ast$ form factor, the pion cloud contribution 
estimated using an S-wave model does not
change with the inclusion of D-waves (44.1\% for model 4 to be compared 
with 46.4\% from Ref.~\cite{NDelta}). 
The magnitude of the cut-off, $\Lambda_\pi^2$, in the pion cloud  
contribution to $G_M^*$ is decreased slightly (1.53 GeV$^2$ 
versus 1.22 GeV$^2$). The addition of the pion clouds term 
to the D-wave states does 
not change the previously observed approximate equality  
$\alpha_1 \simeq \alpha_2$, although its reduces very slightly the
value
of these range parameters.

Even though the pion cloud contributions dominate the description of the 
small form factors, our model suggests that the corrections coming 
from the valence quark sector are still important to 
obtain the best description of the data.
This can be confirmed in Fig.~\ref{fig6}, 
in particular for $R_{SM}$.   Leaving aside the discrepancy 
at high $Q^2$ (which depends on the resolution of 
the differences between the {recent MAID analysis and 
the older JLab analysis) 
the figure shows that the 
pion cloud contribution, which in 
our work is parameter free,
underestimates the data.
% in the region of $Q^2$ from about 0.5 to 2 GeV$^2$.

\begin{figure}[t]
%\vspace{0.0cm}
\centerline{
\mbox{
\includegraphics[width=7cm]{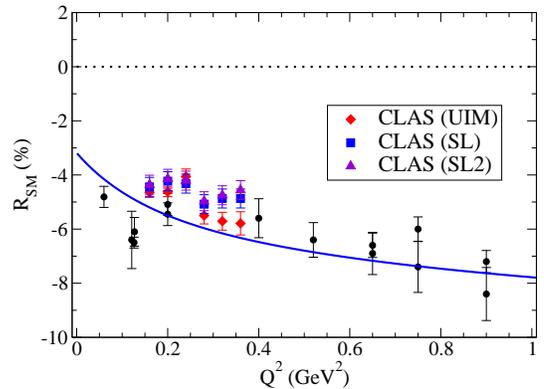}}}
%\vspace{0.9cm}
\caption{$R_{SM}$ from model 4 with data and 
preliminary CLAS data using an Unitary Isobar Model (UIB), 
Sato and Lee model from \cite{SatoLee} (SL) and 
\cite{Diaz06a} (SL2). 
The preliminary CLAS data was not used in the fit.}
\label{figRSMclas}
\end{figure}

While model 4 fits the overall data well, and is clearly the best
model found,
there are at least three ways in which our theory and the experimental data could be improved.
First, the validity of our model for the pion cloud can be questioned, particularly away from the photon point, $Q^2=0$.  
Second, the analysis of the data for $Q^2 < 0.15$  GeV$^2$ 
and for large $Q^2$ is uncertain. 
If the low $Q^2$ data is excluded from the fit, 
we found that a higher quality description
of $R_{SM}$ was  possible,   with  $\chi^2 \sim 1.2$.  
That there is some legitimacy in excluding this data may be seen 
from a comparison between different data sets, shown in Fig.\ \ref{figRSMclas}. 
The experiments for $Q^2=0.121$ GeV$^2$ from MIT-Bates 
and $Q^2= 0.126, \; 0.127$ GeV$^2$ from MAMI
suggest a large negative fraction for the   $R_{SM}$ 
(around -7 \%), in contradiction with 
the data for $Q^2=0.2$ GeV$^2$ from MAMI, and 
also the very recent preliminary CLAS data 
 \cite{Stave08,Diaz06a,Smith07} ($R_{SM} \sim -5 \%$).  The final analysis of the CLAS data should clarify this point.

Finally, our treatment of the valence quark sector can be questioned. 
The factor $f_C$ in Eq.~(\ref{eq:fc}) gives a zero in
$G_C^B$, which implies a dominance of the 
pion cloud for $Q^2 \sim 6$ GeV$^2$.  
%{\bf In the absence of pion cloud effects this behavior 
%is inconsistent with the "bare" quark data.}
To study the model dependence on this behavior of $f_C$,  we probed
a change in the
%we tested  changing the sign of $f_{2-}$,  and reducing the 
quark anomalous magnetic moments $\kappa_\pm$. 
We tried to suppress the large $Q^2$ behavior of $f_{2-}$ by redefining  
$\kappa_\pm \to \kappa_{\pm} \frac{\Lambda^2}{\Lambda^2+Q^2}$,
with $\Lambda$ an adjustable cut-off. This reparameterization 
decreased the overall $\chi^2$  obtained for 
the description of the transition data,
but also increased the $\chi^2$ for the 
fit to the nucleon form factor data.  
Clearly this effect deserves more study.

\section{Comparison with other works }
\label{secResB}

In general our results agree qualitatively 
with  Refs.~\cite{BuchmannEtAl,Buchmann01}
and support the general idea that the 
quadrupole transition form factors are 
dominated by pion cloud effects 
\cite{Giannini07,Bernstein07,Kamalov99,Pascalutsa06b,Burkert04,Stave08}
or quark-antiquark states 
\cite{BuchmannEtAl,Buchmann01,Buchmann04,Buchmann07a}. 
In particular our pion cloud contribution 
is consistent with both chiral perturbation 
and Lattice QCD estimations.
Also, according to results of chiral perturbation theory 
$R_{SM} \sim \log m_\pi$ as $m_\pi \to 0$ 
which implies significant pion cloud effects at the physical pion mass
\cite{Pascalutsa06a,Gail06,Vanderhaeghen07}.
Additionally, effective field theory calculations
of Gail {\it et al.}~\cite{Drechsel06,Gail06} 
predict a dominance of the pion cloud effects.
The recent quenched and unquenched 
lattice QCD calculations  \cite{Alexandrou07}
for pion masses $m_\pi > 0.35$ GeV 
predict only a small fraction of the 
experimental result for $G_C^\ast$ for low $Q^2$.
This fact suggests as well that the pion 
cloud effects are dominant in $G_C^\ast$ 
for low $Q^2$
(the enhancement of the chiral loop 
corrections relative to lattice data for 
small pion masses was shown in Ref.~\cite{Pascalutsa06a}).

%\begin{widetext}

\begin{figure*}
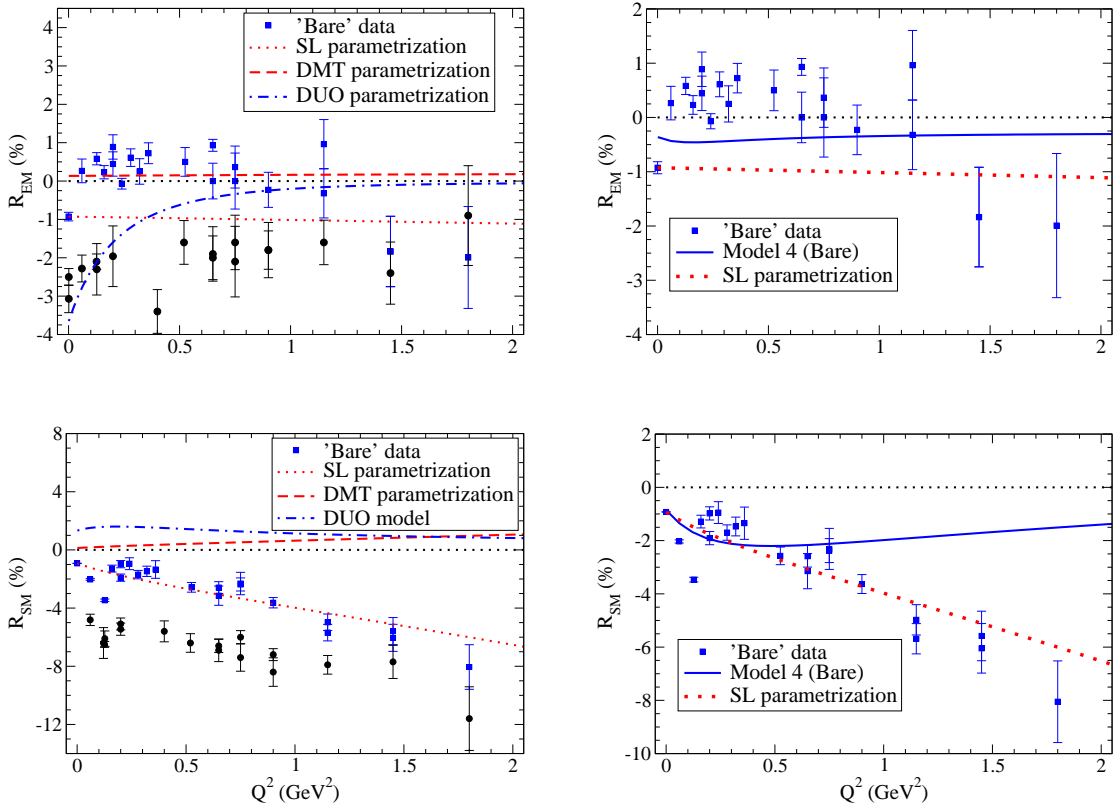

\vspace{1.0cm}
\centerline{
\mbox{
\includegraphics[width=6.9cm]{REMbare2} }  \hspace{.5cm}
\mbox{
\includegraphics[width=6.9cm]{REMmI6b} } }
\vspace{.9cm}
\centerline{
\mbox{
\includegraphics[width=6.9cm]{RSMbare2} }  \hspace{.5cm}
\mbox{
\includegraphics[width=6.9cm]{RSMmI6b} } 
}
%\vspace{-2cm}
\caption{
Left side: 
Parametrization to the ``bare''  form factors from 
Sato and Lee (SL) \cite{SatoLee}, Dubna-Mainz-Taipai (DMT) 
\cite{Kamalov99,Kamalov01} and 
Dynamical Utrecht-Ohio (DUO) \cite{Pascalutsa04} models. 
The circles represents the experimental 
data from Figs.~\ref{fig1} and Fig.~\ref{fig1b}.
Right side: Model 4  compared with SL and ``bare'' data.
The ratios were evaluated using the parametrization 
of $G_M^\ast$ from (\ref{eqGMest}).
In both cases the ``bare'' data is from Ref.~\cite{Diaz06a}.
}
\label{fig3}
\end{figure*}

%\end{widetext}

Constituent quark models, as the
Isgur-Karl model \cite{Pascalutsa06b,Isgur82},
include S-states and a D-state admixture of typically 
1\%, but predict 
only a fraction of the 
total $G_E^\ast$ and $G_C^\ast$ near $Q^2=0$.
This feature is also shared by several 
relativistic quark models \cite{Riska,Capstick90a}
and by the valence contribution of our model 4
(see Figs.~\ref{fig6} and \ref{fig3}).
The exception to this role is the work of Ref.~\cite{Faessler06}
where a manifestly Lorentz covariant 
chiral quark approach was considered. 
In that work the effect of the pion 
cloud is reduced to the order of 10\%, 
which is compensated by a significant 
contribution of relativistic effects
when compared with non-relativistic quark models. 
A discussion of the predictions of quark 
models for low $Q^2$ can be found in 
Refs.~\cite{Giannini07,Pascalutsa06b,Stave06a,Stave08}.

We conclude this section by looking at the implications 
of describing the pion cloud with a dynamical model 
(DM).  Dynamical models assume that the complete electroproduction 
(or photoproduction) amplitude is the iteration 
of a kernel composed of the sum of bare resonance pole(s) 
in the $s$ channel, plus ``left-hand cuts''  
(arising from the angular average of 
$t$ and $u$ channel poles coming from 
the exchanges of mesons and baryons).  
Fitting the data with such a model fixes 
{\it both\/} the bare resonance parameters {\it and\/} 
the parameters of the left-hand cuts 
which dynamically determine the background, 
and in this way allows one to extract  
bare ``data'',  or that part of the 
form factors that would be present even 
without the dressing produced by the 
rescattering of pions.  Predictions of a pure CQM 
could be compared directly with this bare data, 
since both exclude the same physics -- all 
pion rescattering mechanisms.

The bare ``data''  extracted from the fits of  Ref.~\cite{Diaz06a} 
are compared to various theoretical models  in Fig.~\ref{fig3}. 
Since the comparison involves electromagnetic ratios and not absolute quantities, 
we used 
(\ref{eqGMs}) to parametrize $G_M^\ast$ which is present in all the ratios
shown in the figure.  The left hand panels 
show the comparison of bare data to the parametrizations used by Sato and Lee 
(SL) \cite{SatoLee}, Dubna-Mainz-Taipai (DMT) 
\cite{Kamalov99,Kamalov01} and the
Dynamical Utrecht-Ohio (DUO) \cite{Pascalutsa04} models.
Note that there is a substantial 
difference between the ``bare'' data \cite{Diaz06a}
and the parametrization initially used 
in the SL model \cite{SatoLee}, particularly for $R_{EM}$.

The figure eloquently exhibits that 
``bare'' contributions are strongly model dependent,
in their size  and even their sign,  
differing from model to model.
All that we can conclude from these results,
considering also the observed experimental data, 
is that ``bare'' contributions  and 
pion cloud contributions are both sizable (see Table \ref{tabE2C2}).
The exception is the DMT model, 
where the bare contributions are almost 
negligible for low $Q^2$ ($\sim 5\%$).
As for the SL model \cite{SatoLee},  the bare contribution
is 33\% and 36\% for E2 and C2  respectively 
at $Q^2=0$ \cite{Drechsel06}
(corresponding  to 
a pion cloud contribution of 
77\% for E2 and 74\% for C2 at the photon point).
A compilation of the bare 
contribution for E2 and C2 for different models 
is presented in Table  \ref{tabE2C2} 
for $Q^2=0$ and 1 GeV$^2$. 
For a summary of the literature see also 
Refs.~\cite{Pascalutsa06b,Drechsel06,Tiator06}.

%% TABLE III
\begin{table}[t]
\begin{center}
\begin{tabular}{l c c}
 $Q^2=0$ GeV$^2$       & $G_E^B/G_E^\ast$ & $G_C^B/G_C^\ast$ \\
\hline
DMT \cite{Drechsel06}     &  -5.7\%  &  -4.7\%   \\
SL  \cite{Drechsel06}    &  33\%  &   36\%  \\
DUO*                     &   136\%     & -42\%        \\
Buchmann \cite{BuchmannEtAl,Buchmann01}
         & 12\%   &   20\%   \\
Model 4   & 13\%   &   18\%  \\
\hline
 $Q^2=1$ GeV$^2$       & $G_E^B/G_E^\ast$ & $G_C^B/G_C^\ast$ \\
\hline
DMT*                      &  -8.8\%  &  -8.2\%   \\
SL*                       &  56\%    &   51\%  \\
DUO*                      &  11\%    &   -15\%       \\
Model 4   & 17\%   &   18\%  \\
\hline
\end{tabular}
\end{center}
\caption{Bare contribution in different models, estimated by
subtraction of the pion cloud.
In the lines labeled with an ``*'', 
the total result was not available
and  model 4 was used.}
\label{tabE2C2}
\end{table}

\section{Conclusions}
\label{secConclusions}

In this work we introduce for the 
first time the D-states in the covariant 
spectator formalism for the description of
baryons as a quark-diquark systems, 
and apply our formalism to the description of the
form factors of  the  electromagnetic $N \Delta$ transition.
Covariant formalisms provide a
correct treatment of boosts and rotations, which are important to describe 
correctly the kinematics and the dynamics
in  the intermediate $Q^2$ region 
($Q^2 \sim 4$ GeV$^2$).
There are two D-states for the $\Delta$: one for
the valence quark core of spin 3/2 (D3 state), 
the other for valence core of spin 1/2 (D1 state).  
We show that these D states have the 
correct spin structure  in the baryon rest frame.
Within this framework we show here that 
a consistent model,  with 
orthogonal nucleon and $\Delta$ wave functions,
predicts non-vanishing contributions 
for the Electric and Coulomb quadrupole form factors,
an indirect signature of the asymmetry  
of the valence quark distribution in space.

However, we start by finding that the D-states contributions
are not enough to explain the experimental 
data for $G_E^\ast$ and $G_C^\ast$.
An admixture of 8\% D3-state can explain the $R_{EM}$ data, 
but the $R_{SM}$ data
cannot be explained without a D1 component.
Importantly, although, is that
even a very large
admixture of the D1-state cannot explain the high $Q^2$ behavior of this observable.
%is considered,  in which case
%the sucessful descriptin is nevertheless restrict  
%to the interval  ($Q^2 < 1$ GeV$^2$).
This conclusion is consistent with 
results from other constituent quark models 
in the literature.

With this established we had to turn our attention to the pion cloud effects.
We find that the  pion cloud contributions are essential to an accurate 
description of the $\gamma N \to \Delta$ transition, and that our best 
model (model 4) gives a good overall 
description of the $\gamma N \to \Delta$ 
transition form factors.
In this model we used pion cloud effects derived in the large $N_c$ limit, containing
no adjustable parameters, and our fit predicts that the $\Delta$ 
wave function is the sum of a large S-state component 
with an admixture of 0.9\% for the D3-state and reasonable 4.4\% weight
for the D1-state.
The pion cloud dominants $G_E^\ast$ and $G_C^\ast$, with 
contributions of 87\% and 73\% respectively,
at the photon point.
Like the valence quark contribution, 
the pion cloud contribution is also covariant 
because it is based on a covariant description 
of the neutron electric form factor.
As the pion cloud parametrization presented 
here can be justified only for low $Q^2$ 
($Q^2 < 1.5$ GeV$^2$, according with Ref.~\cite{Pascalutsa07a}), 
in the future 
we are planing to include an explicit relativistic
calculation of the pion cloud, to replace 
the effective parametrization.

%In the future we plan to 
%include an explicit calculation of the pion cloud, 
%to replace the effective parametrization.

The momentum distribution 
of the D3 state is determined by one parameter ($\alpha_5\simeq0.20$), 
and that of the D1 state by two parameters which turn out 
to be nearly equal ($\alpha_3\simeq\alpha_4\simeq0.10$).
These values are smaller than the one parameter 
($\alpha_1 \simeq \alpha_2\simeq0.35$) 
required to represent the S state, consistent with 
the picture that the D waves are more peripheral.

We conclude with two notes which concern also future developments:

1)We found that the quality of the description 
of the data is very sensitive in the regions 
$Q^2< 0.2$ GeV$^2$ and at higher $Q^2>3$ GeV$^2$.   
Evidence for problems in the data
come from 
an apparent inconsistency between 
the $G_C^*$ data for $Q^2 \simeq 0.13$ GeV$^2$ and $Q^2 \simeq 0.2$ GeV$^2$.
The new CLAS data  \cite{Stave08,Diaz06a,Smith07}
for 0.16 GeV$^2 \le Q^2 \le$ 0.34 GeV$^2$ 
can be useful to clarify the situation.
For high $Q^2$, the understanding 
of the differences between the CLAS 
analysis and the MAID analysis will 
be also crucial for future progress.

2)Presumably the most accurate estimate of the pion cloud 
effects comes from dynamical models which compute 
the dressing of the bare quark currents by pion 
rescattering to all orders.  
As a byproduct, these dynamical models 
determine the parameters of the valence, 
or undressed, quark contribution, 
which can be compared to a quark model without 
pion cloud effects.   We observed that the 
results from the  bare, pure valence quark form factors 
strongly depend on the  
pion production model (see SL, DMT or DUO models presented in the text).
It becomes therefore crucial to use valence quark 
models to estimate directly the bare form factors 
as functions of $Q^2$, and
to understand the nature of the valence quark distribution 
in the nucleon and $\Delta$ system, as we do here. Since  
the valence quark distribution  dominates 
the largest $\gamma N \to \Delta$ transition form factor, 
$G_M^*$ \cite{NDelta}, dynamical models should use 
valence quark models as input, 
instead of relying on phenomenological parametrizations.
%\cite{Stave08}.

\vspace{0.3cm}
\noindent
{\bf Acknowledgments}

G.~R.\ wants to thank Vadim Guzey,  
Jos\'e Goity and Kazuo Tsushima for the helpful discussions. 
The authors want also to thank 
to B.\ Juli\'a D\'{\i}az  for sharing 
the 'bare data' of Ref.\ \cite{Diaz06a}.
This work was partially support by Jefferson Science Associates, 
LLC under U.S. DOE Contract No. DE-AC05-06OR23177.
G.~R.\ was supported by the portuguese Funda\c{c}\~ao para 
a Ci\^encia e Tecnologia (FCT) under the grant  
SFRH/BPD/26886/2006.

\appendix

%%\renewcommand{\theequation}{A.\arabic{equation}}
% redefine the command that creates the equation no.
%%\setcounter{equation}{0}  % reset counter

\section{Spin structure of the D-wave matrix elements} \label{apDwave}

In this appendix we first show how the spin 3/2 function $V_{3/2}$ 
[defined in Eq.~(\ref{eq:233})] satisfies the special spin 3/2 constraint
\bea
\gamma_\alpha\,V^\alpha_{3/2}(P,\lambda)=0
\eea
 and then prove the relations (\ref{eq:234}) and explicitly construct  
the matrix elements in the expansions  (\ref{eq:D1&3}).

To show that $V_{3/2}$ satisfies the special spin 3/2 condition, 
go to the rest frame $\overline{P}=(M,0,0,0)$ and observe that 
\bea
 \gamma_\alpha\, V^\alpha_S(\overline{P},\lambda_s)=\sum_\lambda\left(\begin{array}{cc} 0 & -a_{\lambda_s \lambda} \cr a_{\lambda_s \lambda} & 0\end{array}\right) \left[\begin{array}{c} \chi_\lambda  \cr  0\end{array}\right] ,
\label{eqaux2}
\eea
where the 2$\times$2 operator is
\bea
a_{\lambda_s \lambda}= \left<\sfrac12 \lambda\; 1 \lambda' 
|\sfrac32 \lambda_s \right> \sigma_{\lambda'}
\eea
with
\bea
\sigma_\lambda=\sigma_i\varepsilon^i_\lambda=\begin{cases} \qquad\sigma_z & \lambda=0 \cr 
\left(\begin{array}{lr} 0 & -\sqrt{2}\cr 0 & 0\end{array}\right) & \lambda=1\cr 
%\begin{array}{lr}           &                &   \end{array}       &          \cr
\left(\begin{array}{lr} 0 &0\cr  \sqrt{2} &\phantom{-} 0\end{array}\right) & \lambda=-1
\end{cases}
\eea
Examining the $\lambda_s>0$ cases shows that
\bea
&&\sum_\lambda \left(a_{\frac32\; \lambda}\right)\chi_\lambda=\sigma_1\chi_+=0\nonumber\\
&&\sum_\lambda \left(a_{\frac12\; \lambda}\right)\chi_\lambda=\sqrt{\sfrac23}\sigma_0\chi_++\sqrt{\sfrac13}\sigma_1\chi_-=0\, .\qquad
\eea
Similar results hold for the $\lambda_s<0$ cases.

The orthogonality and normalization relations (\ref{eq:234}) 
follow immediately from the orthogonality and normalization 
of the spinors and polarization vectors, and the unitarity 
of the CG coefficient matrix.  The completeness relations 
follow from the normalization and orthogonality relations, 
but it is instructive to prove them directly.   
To do this go to the rest frame and compute the 
matrix elements of the projectors in the spherical basis.  Define
\bea
P^S_{\lambda\lambda'}=\varepsilon_{\lambda\,P}^{\alpha\,*}\,({\cal P}_S)_{\alpha\beta}\,\varepsilon^\beta_{\lambda'\,P}
\eea
and note that this matrix is hermitian and when multiplied by the projection operator $(M+\overline{\not\!P})/(2 M)=\frac12(1+\gamma^0)$ is of the block diagonal form
\bea
P^S_{\lambda\lambda'}=\left(\begin{array}{cc} -p^S_{\lambda\lambda'}& 0 \cr 0 & 0 
%p^S_{\lambda\lambda'}
\end{array}\right)
\eea
where the $2\times2$ submatrices have the symmetry property
\bea
\sigma_x p^S_{\lambda\lambda'}\,\sigma_x= p^S_{-\!\lambda\,-\!\lambda'}\, .
\eea
Hence there are only three independent elements, which will be chosen to be $p_{00}$, $p_{11}$, and $p_{01}$.  Using the definitions of the projection operators (\ref{eqP32}),  in the rest frame we obtain directly
\bea
&p^{1/2}_{00}=\frac13\left[\begin{array}{cc} 1&0\cr 0 & 1\end{array}\right] &p^{3/2}_{00}=\sfrac23\left[\begin{array}{cc} 1&0\cr 0 & 1\end{array}\right]\nonumber\\
&p^{1/2}_{11}=\frac13\left[\begin{array}{cc} 0&0\cr 0 & 2\end{array}\right] &p^{3/2}_{11}=\sfrac13\left[\begin{array}{cc} 3&0\cr 0 & 1\end{array}\right]\nonumber\\
&p^{1/2}_{10}=-\frac13\left[\begin{array}{cc} 0&\sqrt{2}\cr 0 & 0\end{array}\right]\quad &p^{3/2}_{10}=\sfrac13\left[\begin{array}{cc} 0&\sqrt{2}\cr 0 & 0\end{array}\right]\, . \label{eq:a9}
\eea

These same operators can be calculated from the definitions (\ref{eq:233}) and the expansions (\ref{eq:234}).  In the rest system the spherical components of these expansions are
\begin{align}
{\cal O}^S_{\lambda\lambda'}\equiv\sum_{\lambda_s} \varepsilon_{\lambda\,\overline{P}}^{\alpha\,*}\,&V_{S\,\alpha}(\overline{P},\lambda_s) \overline{V}^\beta_S(\overline{P},\lambda_s)\,\varepsilon_{\lambda'\,\overline{P}}^{\beta}
\nonumber\\
&=\sum_{\lambda_s} \left<\sfrac12 \,\lambda_1 \;1\,\lambda |S\,\lambda_s\right>\left<\sfrac12 \,\lambda'_1 \;1\,\lambda' |S\,\lambda_s\right>
\nonumber\\
&\qquad\qquad\times u_\Delta(\overline{P},\lambda_1)\overline{u}_\Delta(\overline{P},\lambda'_1)
\end{align}
where the operator ${\cal O}$ has the form
\bea
{\cal O}^S_{\lambda\lambda'}=\left(\begin{array}{cc} -o^S_{\lambda\lambda'}&0\cr 0 & 0\end{array}\right)
\eea
with the 2$\times$2 matrix
\bea
o^S_{\lambda\lambda'}=\sum_{\lambda_s} \left<\sfrac12 \,\lambda_1 \;1\,\lambda |S\,\lambda_s\right>\left<\sfrac12 \,\lambda'_1 \;1\,\lambda' |S\,\lambda_s\right>\chi_{\lambda_1}\chi^\dagger_{\lambda'_1}\, .\qquad
\eea
This operator has the same symmetry properties as $p^S_{\lambda\lambda'}$, and by explicit computation 
\bea
o^S_{00}&=&\Big[\left<\sfrac12 \,\sfrac12 \;1\,0 |S\,\sfrac12 \right>\Big]^2\left(\begin{array}{cc} 1&0\cr 0 & 0\end{array}\right) 
\nonumber\\
&&+ \Big[\left<\sfrac12 \,-\!\sfrac12 \;1\,0 |S\,-\!\sfrac12 \right>\Big]^2\left(\begin{array}{cc} 0&0\cr 0 & 1\end{array}\right) \nonumber\\
&=&\begin{cases}  \frac13 \left(\begin{array}{cc} 1&0\cr 0 & 1\end{array}\right) & S=\frac12 \cr \frac23 \left(\begin{array}{cc} 1&0\cr 0 & 1\end{array}\right) & S=\frac32\end{cases}
\eea
in agreement with Eq.~(\ref{eq:a9}).  Similar agreement is obtained for the other matrices. 

We now turn to the computation of the matrix elements in Eq.~(\ref{eq:D1&3}).  Using the direct product representations (\ref{eq:233}) (and a similar one for $w_\gamma$)
\begin{align}
C_{D\,2S}&\equiv\overline{V}_{S\,\alpha}(P,\lambda_s){\cal D}^{\alpha\gamma}w_\gamma(P,\lambda_\Delta)
\nonumber\\
&=\sum_\lambda\left<\sfrac12 \,\lambda \;1\,m |S\,\lambda_s\right>\left<\sfrac12 \,\lambda \;1\,m' |\sfrac32\,\lambda_\Delta\right>D_{m m'}\nonumber\\
&=\sfrac13\sqrt{8\pi}\,{\bf k}^2 Y^2_{m_\ell}\sum_\lambda\left<2 \,m_\ell\;1\,m |1\,m'\right>
\nonumber\\
&\qquad\quad\times\left<1\,m' \;\sfrac12 \,\lambda|\sfrac32\,\lambda_\Delta\right> \left<1\,m\;\sfrac12 \,\lambda |S\,\lambda_s\right>,
\end{align}
where the CG coefficients guarantee that $m=\lambda_s-\lambda$, 
$m'=\lambda_\Delta-\lambda$, and 
$m_\ell=m'-m= \lambda_\Delta-\lambda_s$.
The sum over three CG coefficients is evaluated using Racah coefficients $W$.  
For the cases at hand \cite{newreference}: 
\begin{align}
&\sum_\lambda\left<2 \,m_\ell\;1\,m |1\,m'\right>
\left<1\,m' \;\sfrac12 \,\lambda|\sfrac32\,\lambda_\Delta\right> \left<1\,m\;\sfrac12 \,\lambda |S\,\lambda_s\right>\nonumber\\
&=\sqrt{3(2S+1)} \,W(2,1,\sfrac32,\sfrac12;1,S)
\left<2 \,m_\ell\;S\,\lambda_s |\sfrac32\,\lambda_\Delta\right>\nonumber\\
&=-(-1)^{\sfrac12-S}\frac{1}{\sqrt{2}}\left<2 \,m_\ell\;S\,\lambda_s |\sfrac32\,\lambda_\Delta\right>,
\end{align}
giving Eq.~(\ref{eq:D1&3}).

\section{Integration in $k$}
\label{apIntK}

When the currents associated to the $\Delta$ D-states 
are written there is a dependence in the 
tensor 
\be
I^{\alpha \beta}(P_+,P_-)=
\int_k {\cal D}^{\alpha \beta}(P_+,k) \psi_\Delta^{D \; 2S} (P_+,k) 
\psi_{N}^S (P_-,k). 
\label{eqIdef}
\ee
The properties of $I^{\alpha \beta}$ in a Lorentz 
transformation follows the properties of ${\cal D}^{\alpha \beta}$:
\be
{\cal D}^{\alpha \beta}(P_+^\prime, k^\prime)=
\Lambda^\alpha_{\; \sigma} 
\Lambda^\beta_{\; \rho}  {\cal D}^{\sigma \rho}(P_+, k).
\ee
Then 
\be
I^{\alpha \beta}(P_+^\prime, P_-^\prime)=
\Lambda^\alpha_{\; \sigma} 
\Lambda^\beta_{\; \rho}  I^{\sigma \rho}(P_+, P_-).
\label{eqItrans}
\ee
In these condition a covariant expression for $I^{\alpha \beta}(P_+,P_-)$
can be derived in a particular frame and the extended for 
an arbitrary frame using (\ref{eqItrans}).

Consider Eq.\ (\ref{eqIdef}) in the $\Delta$ rest frame.
In the $\Delta$ rest frame the scalar wave functions 
are independent of the variable $\varphi$.
In these conditions 
we can perform the analytical 
integration in $\varphi$, replacing the integral expression,  
by a equivalent integral with an integrand function 
independent of $\varphi$.
The result is
\be
\frac{1}{2\pi}
\int d \varphi {\cal D}^{\alpha \beta}(P_+,k)=  
S_3 \bar R^{\alpha \beta},
\label{eqIntPHI}
\ee
where $z=\cos \theta$ ($\theta$ is the angle 
between ${\bf k}$ and ${\bf q}$) and
\ba
S_3&=& \frac{{\bf k}^2}{2}
(1-  3  z^2)
%\nonumber 
\label{eqS3}
\\
\bar R^{\alpha \beta} &=& 
\left[ \begin{array}{cccc}
0 & 0 & 0 & 0  \\
0  & 1/3 & 0 & 0 \\
0  & 0  & 1/3  & 0 \\
0  & 0  & 0  & -2/3 \\
\end{array} \right].
\label{eqRbar}
\ea
We can express (\ref{eqIntPHI}) in a covariant 
form considering covariant expressions for $S_3$ and 
$\bar R^{\alpha \beta}$ 
\ba
& & S_3 \to b(k,q) \nonumber \\
& & \bar R^{\alpha \beta} \to 
R^{\alpha \beta} (P_+,P_-).
\nonumber 
\ea 
In particular we can write in the $\Delta$ 
rest frame
\ba
& &
b(\tilde k, \tilde q) =
\frac{3}{2} \frac{(\tilde k \cdot \tilde q)^2}
{\tilde q^2}
-\frac{1}{2} \tilde k^2 
\label{eqBtilde}
\\
& &R^{\alpha \beta} (P_+,P_-)
=
\frac{\tilde q^\alpha  
\tilde q^\beta}{\tilde q^{2}}
-\frac{1}{3} \tilde g^{\alpha \beta}.
\label{eqRtilde}
\ea

The expression $R^{\alpha \beta} (P_+,P_-)$ is 
the only covariant expression for $\bar R^{\alpha \beta}$
compatible with (\ref{eqRbar}), as can be showed 
considering the most general expression:
\ba
& &
R^{\alpha \beta}(P_+,P_-)=
\frac{a}{M^2} P_+^\alpha P_+^\beta+
\frac{b}{M^2} P_+^\alpha P_-^\beta \nonumber \\
& &
+\frac{c}{M^2} P_-^\alpha P_+^\beta+
\frac{d}{M^2} P_-^\alpha P_-^\beta+
e\;g^{\alpha \beta},
\label{eqRgen}
\ea
where $a$, $b$, $c$, $d$ and $e$ 
are functions of $Q^2$.

Similarly, the identity (\ref{eqBtilde}) is the only 
possible covariant representation of $S_3$.
Equivalent representation involving 
the factors $\tilde q \cdot k$
or $q \cdot \tilde k$ are reduced to $\tilde q \cdot \tilde k$.
By definition of $\tilde k$ and $\tilde q$: 
$\tilde k \cdot q= \tilde q \cdot k  =\tilde q \cdot \tilde k$.

As consequence of the representation (\ref{eqBtilde}) and 
(\ref{eqRtilde}), the integral (\ref{eqIntPHI})
in the $\Delta$ rest frame can be 
represented by 
\be
\frac{1}{2\pi}
\int d \varphi
%\int_\varphi 
{\cal D}^{\alpha \beta}(P_+,k)=  
b(\tilde k,\tilde q) R^{\alpha \beta }(P_+,P_-).
\label{eqIntPHI2}
\ee
As (\ref{eqIntPHI2}) is expressed in a covariant 
notation we can obtain the equivalent 
expression for a different collinear frame 
considering an appropriate boost.

Using Eq.\ (\ref{eqIntPHI2}) we can write 
for any collinear frame 
\be
I^{\alpha \beta }(P_+,P_-)=
R^{\alpha \beta }(P_+,P_-)
I_D(P_+,P_-),
%\label{eqID0}
\ee
where
\ba
& &
I_D(P_+,P_-)=
\int_k b(\tilde k,\tilde q) 
\psi_\Delta^{D \; 2S } (P_+ ,k) 
\psi_N^S(P_-,k). \nonumber 
\\
%\label{eqIDp}
\ea

%
%In the last equation $\psi_D$ holds for any $\Delta$ D-state.

%\input{biblo.tex}

%%%%%%%%%%%%%%%%%%%%%%%%%%%%%%%%%%%%%%%%%%%%%%%%%%%

\end{document}